\documentclass[prodmode,acmtomccap]{acmlarge}

\usepackage[T1]{fontenc}
\usepackage[latin9]{inputenc}
\usepackage[english]{babel}
\usepackage{verbatim}
\usepackage{float}
\usepackage{color}
\usepackage{amsmath}
\usepackage{amssymb}
\usepackage{graphicx}
\usepackage{esint}
\usepackage[unicode=true,
 bookmarks=true,bookmarksnumbered=false,bookmarksopen=false,
 breaklinks=false,pdfborder={0 0 0},backref=false,colorlinks=false]
 {hyperref}

\makeatletter

\providecommand{\tabularnewline}{\\}
\floatstyle{ruled}

\usepackage{epsfig}
\usepackage{xspace}
\usepackage{algorithmic}
\DeclareMathOperator{\tr}{tr}
\pdfpageattr {/Group << /S /Transparency /I true /CS /DeviceRGB>>} 

\newcommand{\pb}{\ensuremath{P_{B}}\xspace}  
\newcommand{\pc}{\ensuremath{P_{C}}\xspace}
\newcommand{\bb}{\mathbb}
\newcommand{\bs}{\boldsymbol}
\newcommand{\mbf}{\mathbf}

\@ifundefined{showcaptionsetup}{}{%
 \PassOptionsToPackage{caption=false}{subfig}}
\usepackage{subfig}
\makeatother

\acmVolume{2}
\acmNumber{3}
\acmArticle{1}
\articleSeq{1}
\acmYear{2013}
\acmMonth{10}

\usepackage[ruled]{algorithm2e}
\SetAlFnt{\algofont}
\SetAlCapFnt{\algofont}
\SetAlCapNameFnt{\algofont}
\SetAlCapHSkip{0pt}
\IncMargin{-\parindent}
\renewcommand{\algorithmcfname}{ALGORITHM}

\title{Exploration in Interactive Personalized Music Recommendation: A
  Reinforcement Learning Approach} \author{XINXI WANG \affil{National
    University of Singapore}
  YI WANG \affil{Institute of High Performance Computing, A*STAR}
  DAVID HSU \affil{National University of Singapore} 
  YE WANG \affil{National University of Singapore}}

\begin{abstract}
  Current music recommender systems typically act in a greedy fashion
  by recommending songs with the highest user ratings. Greedy
  recommendation, however, is suboptimal over the long term: it does
  not actively gather information on user preferences and fails to
  recommend \textit{novel} songs that are potentially interesting. A
  successful recommender system must balance the needs to
  \textit{explore} user preferences and to \textit{exploit} this
  information for recommendation. This paper presents a new approach
  to music recommendation by formulating this exploration-exploitation
  trade-off as a reinforcement learning task called the multi-armed
  bandit. To learn user preferences, it uses a Bayesian model, which
  accounts for both audio content and the novelty of recommendations.
  A piecewise-linear approximation to the model and a variational
  inference algorithm are employed to speed up Bayesian inference. One
  additional benefit of our approach is a single unified model for
  both music recommendation and playlist generation. Both simulation
  results and a user study indicate strong potential for the new
  approach.
\end{abstract}

\category{H.3.3}{Information Search and Retrieval}{Information
  filtering} \category{H.5.5}{Sound and Music Computing}{Modeling,
  Signal analysis, synthesis and processing}

\terms{Algorithm, Model, Application, Experimentation}
\keywords{Recommender systems, Machine Learning, Music}

\acmformat{Wang, X., Wang, Y., Hsu, D., Wang, Y. 2013. Exploration in
  Interactive Personalized Music Recommendation: A Reinforcement
  Learning Approach.}

\begin{document}

\begin{bottomstuff}
  This research is supported by the Singapore National Research
  Foundation under its International Research Centre @ Singapore
  Funding Initiative
  and administered by the IDM Programme Office. \\
  Author's address: X. Wang, D. Hsu and Ye Wang are with the
  department of Computer Science, National University of Singapore,
  SG, 117417, e-mail: \{wangxinxi,wangye,dyhsu\}@comp.nus.edu.sg; Yi
  Wang is with the Computing Science Department at IHPC, A*STAR, SG,
  138632, e-mail: wangyi@ihpc.a-star.edu.sg.
\end{bottomstuff}
\maketitle

\section{Introduction}
\label{sec:Introduction}
A music recommendation system recommends songs from a large database
by matching songs with a user's preferences. An \textit{interactive}
recommender system infers the user's preferences by incorporating user
feedback into recommendations. Each recommendation thus serves two
objectives: (i) satisfy the user's current musical need, and (ii)
elicit user feedback in order to improve future recommendations.
\begin{figure}
\begin{centering}
\vspace{-10mm}
\begin{tabular*}{0.8\linewidth}{@{\extracolsep{\fill}}ccc}
\begin{tabular}{c}
  \subfloat[Rating table\label{fig:Rating-matrix}]{\begin{centering}
      \includegraphics[bb=19bp 0bp 78bp 70bp,scale=1.05]{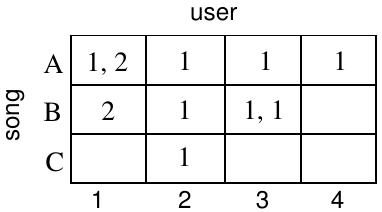}\vspace{10mm}
    \end{centering}
} \vspace{5mm}\tabularnewline
\subfloat[Predicted rating distributions\label{fig:Predicted-rating-distribution}]{\begin{centering}
  \includegraphics[bb=0bp 25bp 230bp 140bp,scale=0.55]{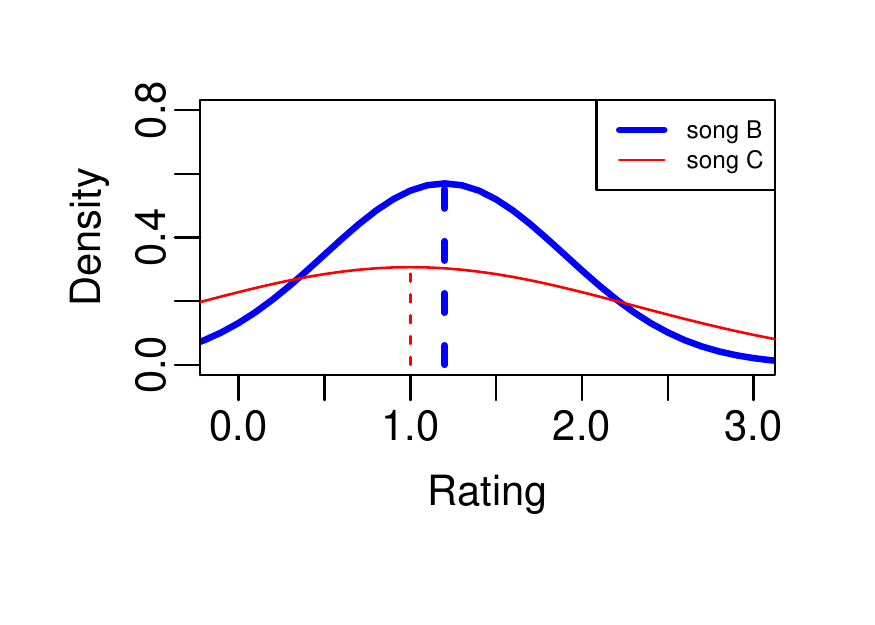}
  \end{centering}}
\end{tabular} & 
\hspace{8mm}\subfloat[Refining the predictions using \newline the greedy strategy (pure exploitation)\label{fig:Greedy}]{%
\vspace{-20mm}
\begin{tabular}{c}
\includegraphics[bb=20bp 32bp 230bp 180bp,clip,scale=0.55]{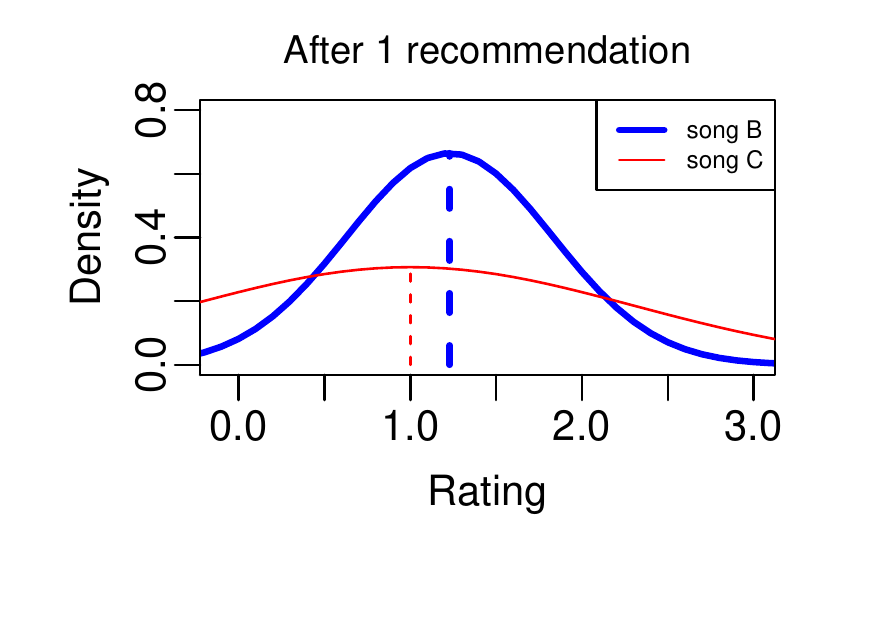}\tabularnewline
\includegraphics[bb=20bp 32bp 230bp 180bp,clip,scale=0.55]{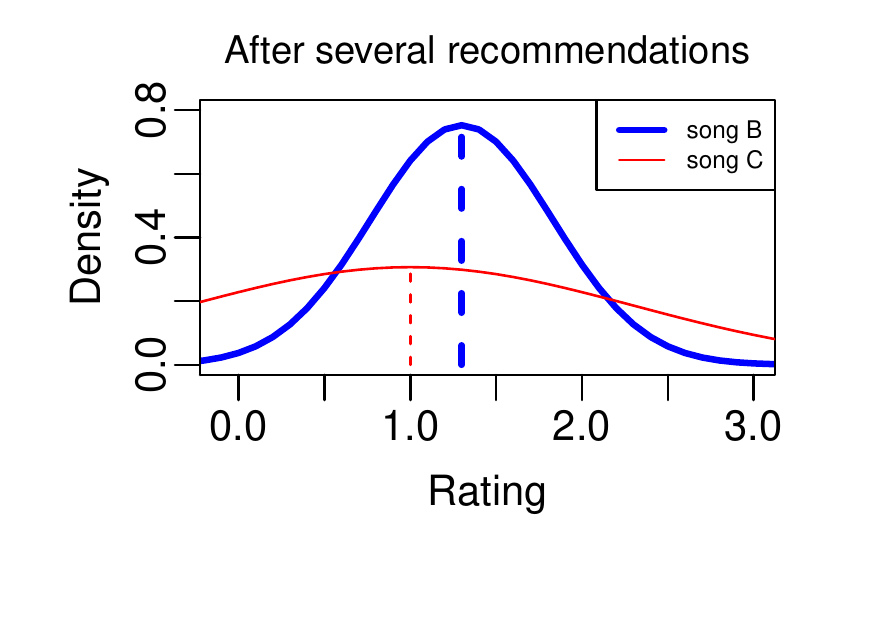}\tabularnewline
\end{tabular}}\hspace{-5mm} & 

\hspace{8mm}
\subfloat[Refining the predictions using multi-armed bandit (exploration/exploitation tradeoff)\label{fig:Multi-armed-bandit}]{%
\vspace{-20mm}
\begin{tabular}{c}
\includegraphics[bb=20bp 32bp 230bp 180bp,clip,scale=0.55]{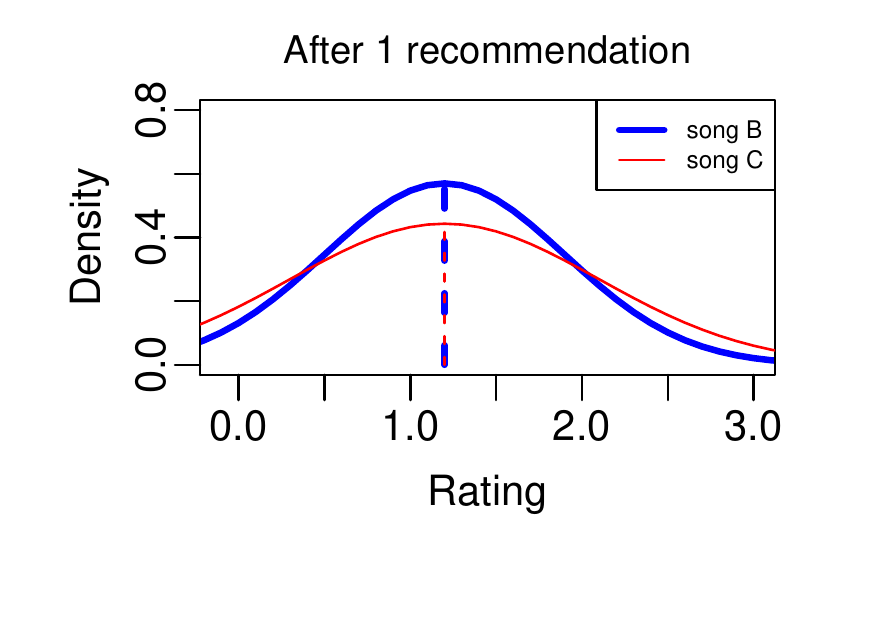}\tabularnewline
\includegraphics[bb=20bp 32bp 230bp 180bp,clip,scale=0.55]{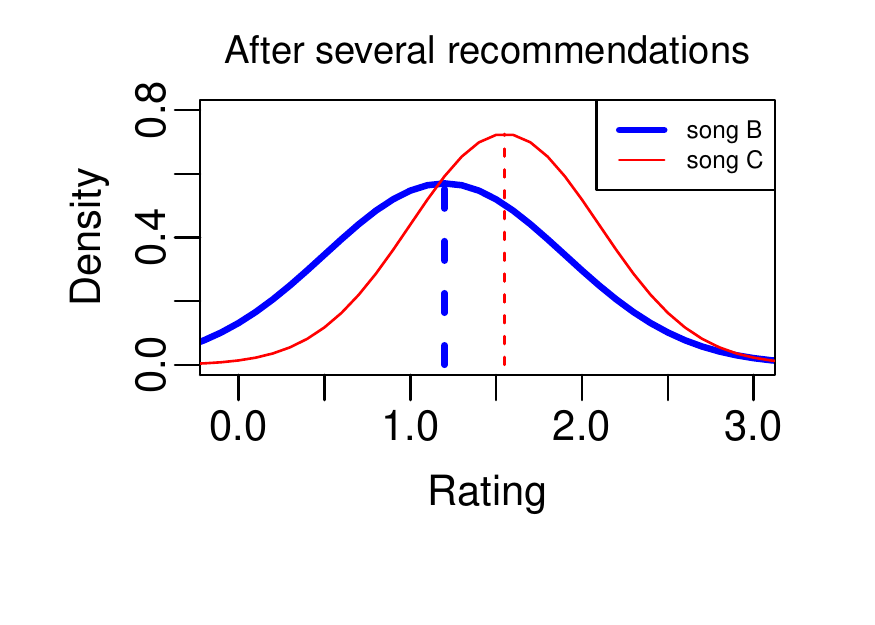}\tabularnewline
\end{tabular}

}
\end{tabular*} 
\end{centering}

\caption{Uncertainty in recommendation\label{fig:Uncertainty-in-recommendation.}}
\end{figure}

Current recommender systems typically focus on the first objective,
while completely ignoring the other. They recommend songs with the
highest user ratings. Such a greedy strategy, which does not actively
seek user feedback, often results in suboptimal recommendations over
the long term.

Consider the simple example in
Figure~\ref{fig:Uncertainty-in-recommendation.}.  The table contains
the ratings for three songs by four users
(Figure~\ref{fig:Rating-matrix}).  For simplicity, let us assume that
that the recommender chooses between two songs $B$ and $C$ only. The
target user is 4, whose true ratings for $B$ and $C$ are $1.3$ and
$1.6$, respectively. The true ratings are real numbers, because a user
may give the same song different ratings as a result of external
factors. The true rating is the expected rating of a song by the
user. In this case, a good recommender should choose $C$.

Since the true user ratings are unknown to the recommender, it may
approximate the rating distributions for $B$ and $C$ as Gaussians, \pb
and \pc (Figure~\ref{fig:Predicted-rating-distribution}),
respectively, using the data in Figure~\ref{fig:Rating-matrix}.  The
distribution \pb has mean $1.2$. The distribution \pc has mean~$1$.
\pb has much lower variance than \pc, because $B$ has more rating
data. A greedy recommender, including, e.g., the highly successful
collaborative filtering (CF) approach, recommends the song with the
highest mean rating and thus~$B$. In response to this recommendation,
user 4 gives a rating, whose expected value is $1.3$. The net effect
is that the mean of \pb likely shifts towards $1.3$ and its variance
further reduces (Figure~\ref{fig:Greedy}). Consequently the greedy
recommender is even more convinced that user 4 favors $B$ and will
always choose $B$ for all future recommendations. It will never choose
$C$ and find out its true rating, resulting in clearly suboptimal
performance.

To overcome this difficulty, the recommender must take into account
\textit{uncertainty} in the mean ratings. If it considers both the
mean and the variance of the rating distribution, the recommendation
will change. Consider again
Figure~\ref{fig:Predicted-rating-distribution}.  Although \pc has
slightly lower mean than \pb, it has very high variance. It may be
worthwhile to recommend it and gather additional user feedback in
order to reduce the variance. User~4's rating on $C$ has expected
value $1.6$. Therefore, after one recommendation, the mean of \pc will
likely shift towards $1.6$ (Figure~\ref{fig:Multi-armed-bandit}).  By
recommending $C$ several times and gathering user feedback, we will
then find out user 4's true preference $C$.

The gist here is that a good interactive music recommender system
must \textit{explore} user preferences actively rather than merely
\textit{exploit} rating information available. Balancing exploration
and exploitation is critical, especially when the system is faced
with a \textit{cold start}, i.e., when a new user or a new song appears.

Another crucial issue for music recommendation is playlist generation.
People often listen to a group of related songs together and may
repeat the same song multiple times. This is unique to music
recommendation and does not occur often for other recommendation
domains, such as newspaper articles or movies. A playlist is a group
of songs arranged in a suitable order. The songs in a playlist have
strong interdependencies.  For example, they share the same
genre~\cite{Chen2012Playlist}, but are diversified at the same
time~\cite{Zhang2012Auralist}. They have a consistent
mood~\cite{Logan2002ContentBased}. They may repeat, but are not
repetitive. Existing recommender systems based on CF or audio content
analysis typically recommend one song at a time and do not consider
their interdependencies during the recommendation process. They divide
playlist generation into two distinct steps~\cite{Chen2012Playlist}.
First, choose a set of favored songs through CF or content analysis.
Next, arrange the songs into a suitable order in a process called
automatic playlist generation (APG).

In this work, we formulate interactive, personalized music
recommendation as a reinforcement learning task called the
\textit{multi-armed bandit}~\cite{Sutton1998Reinforcement} and address
both exploration-exploitation trade-off and that of playlist
generation with a single unified model:
\begin{itemize}
\item Our bandit approach systematically balances exploration and
  exploitation, a central issue well studied in reinforcement
  learning. Experimental results show that our recommender system
  mitigates the difficulty of cold start and improves recommendation
  performance, compared with the traditional greedy approach.
\item We build a single rating model that captures both user
  preference over audio content and the novelty of recommendations. It
  seamlessly integrates music recommendation and playlist generation.
\item We also present an approximation to the rating model and new
  probabilistic inference algorithms in order to achieve real-time
  recommendation performance.
\end{itemize}

Although our approach is designed specifically for music
recommendation, it is possible to be generalized to other media types
as well. The detailed discussion will be presented in
Section~\ref{discussion}.

In the following, Section~\ref{sec:Related-Work} describes related
work. Section~\ref{sec:Modeling} formulates the rating model and our
bandit approach to music
recommendation. Section~\ref{sec:Bayesian-Regression-Models-And-Inference}
presents the approximate Bayesian models and inference algorithms.
Section~\ref{sec:Experiments} describes the evaluations of our models
and algorithms. Section~\ref{discussion} discusses the possible
generalization directions of the approach and future research
directions. Section~\ref{sec:Discussion-and-future} concludes this
work.

\section{Related Work }

\label{sec:Related-Work}

\subsection{Music recommendation}

Since \cite{Song2012Survey} have given a very recent and comprehensive
review of existing music recommendation works, we will first provide a
brief overview of the status quo and discuss highly relevant work in
detail later. Currently, music recommender systems can be classified
according to their methodologies into four categories:
\emph{collaborative filtering} (CF), \emph{content-based} methods,
\emph{context-based} methods, and hybrid methods. Collaborative
filtering recommends songs by considering those preferred by other
like-minded users. The state-of-the-art method for performing CF is
non-negative matrix factorization, which is well summarized
by~\cite{Koren2009Matrix}. Although CF is one of the most widely used
methods, it suffers from the notorious cold-start problem since it
cannot recommend songs to new users whose preference are unknown (the
\textit{new-user} problem) or recommend new songs to users (the
\textit{new-song} problem). Unlike CF, content-based method recommends
songs which have similar audio content to the user's preferred
songs. The recommendation quality of content-based systems is largely
determined by acoustic features, the most useful ones of which, timbre
and rhythm, are incorporated into our proposed
system~\cite{Song2012Survey}.  Content-based systems remedies the
new-song problem but not the new-user problem. Recently, context-based
music recommender systems have become popular. They recommend songs to
match various aspects of the user context, e.g., activities,
environment, mood, physiological states~\cite{Wang2012Context}.
Hybrid methods combine two or more of the above methods.

Relatively few works have attempted to combine music recommendation
with playlist generation. In \cite{Chen2012Playlist}, a playlist is
modeled as a Markov process whose transition probability models both
user preferences and playlist coherence. In
~\cite{Zheleva2010Statistical}, a model similar to Latent Dirichlet
Allocation is used to capture user latent taste and mood of
songs. In~\cite{Aizenberg2012Build}, a new CF model is developed to
model playlists in Internet radio stations. While the three works also
combine recommendation with playlist generation, our model differs in
three aspects: (1) it is based on audio content while the previous
three depend only on usage data; (2) our model is highly efficient so
allowing easy online updates; (3) our model is crafted and evaluated
based on real-life user interaction data, not data crawled from the
web. Zhang \textit{et al.}  tries to recommend using a linear
combination of CF's results with the results from an existing novelty
model~\cite{Zhang2012Auralist}, which ranks songs by CF before
generating the playlists according to novelty. The parameters for the
linear combination are adjusted manually, not optimized
simultaneously.  Moreover, they provide only system\nobreakdash-wise
control of novelty while our method provides user\nobreakdash-wise
control. Other works like~\cite{Hu2011NEXTONE} generate music
playlists within a user's own music library, in which case his/her
preference is already known and need not to be inferred.

\subsection{Reinforcement learning}

\label{sec:background}Unlike supervised learning
(e.g. classification), which considers only prescribed training data,
a reinforcement learning (RL) algorithm \emph{actively} explores its
environment to gather information and exploits the learnt knowledge to
make decision or prediction.

Multi-armed bandit is the most thoroughly studied reinforcement
learning problem. For a bandit (slot) machine with $M$ arms, pulling
arm $i$ will result in a random payoff $r$, sampled from an unknown
and arm-specific distribution $p_{i}$. The objective is to maximize
the \textit{total payoff} given a number of interactions. Namely, the
set of arms is $\mathcal{A}=\{1\dots M\}$, known to the player; each
arm $i\in\mathcal{A}$ has a probability distribution $p_{i}$, unknown
to the player. The player also knows he has $n$ rounds of pulls. At
the $l$-th round, he can pull an arm $I_{l}\in\mathcal{A}$, and
receive a random payoff $r_{I_{l}}$, sampled from the distribution
$p_{I_{l}}$. The objective is to wisely choose the $n$ arms, i.e.,
$(I_{1},I_{2},\dots I_{n})\in\mathcal{A}^{n}$ to maximize
\[
\mbox{Total payoff}=\sum_{l=1}^{n}r_{I_{l}}
\]

A naive solution to the problem could be: the player first randomly
pulls arms to gather information to learn $p_{i}$ (exploration) and
then always pulls the arm that yields the maximum predicted payoff
(exploitation). However, both too much exploration, i.e., the learnt
information is not used much, or too much exploitation, i.e., the
player lacks information to make accurate predictions, result in
suboptimal total payoff. Thus, how to balance the amount of the two is
important.

Multi-armed bandit approach provides a principled solution to this
problem. The simplest multi-armed bandit approach, namely
$\epsilon$-greedy, chooses the arm with the highest predicted payoff
with probability $1-\epsilon$ or chooses arms uniformly at random with
probability $\epsilon$. An approach better than $\epsilon$-greedy is
based on a simple and elegant idea called upper confidence bound
(UCB)~\cite{Auer2002Using}.  Let $U_{i}$ be the true expected payoff
for arm $i$, i.e., the expectation of $p_{i}$; UCB-based algorithms
estimate both its expected payoff $\hat{U}_{i}$ and a confidence bound
$c_{i}$ from history payoffs, so that $U_{i}$ lies in
$(\hat{U}_{i}-c_{i},\hat{U}_{i}+c_{i})$ with high
probability. Intuitively, selecting an arm with large $\hat{U}_{i}$
corresponds to exploitation, while selecting one with large $c_{i}$
corresponds to exploration. To balance exploration and exploitation,
UCB-based algorithms always select the arm that maximizes
$\hat{U}_{i}+c_{i}$, the principle of which is called {}``optimism in
the face of uncertainty''.

Bayes-UCB~\cite{KaufmannCG12} is one of the state-of-the-art Bayesian
counterparts of the UCB approach. In Bayes-UCB, the expected payoff
$U_{i}$ is regarded as a random variable, and the posterior
distribution of $U_{i}$ given the history payoffs $\mathcal{D}$,
denoted as $p(U_{i}|\mathcal{D})$, is maintained, and the fixed-level
\textit{quantile} of $p(U_{i}|\mathcal{D})$ is used to mimic the upper
confidence bound. Similar to UCB, every time Bayes-UCB selects the arm
with the maximum quantile. More interestingly, UCB-based algorithms
require an explicit form of the confidence bound, which is difficult
to derive in our case, but in Bayes-UCB, the quantiles of the
posterior distributions of $U_{i}$ can be easily obtained using
\textit{Bayesian inference}. We therefore choose Bayes-UCB.

There are more sophisticated RL methods such as Markov Decision
Process (MDP) \cite{szepesvari:reinforcement}, which generalizes the
bandit problem by assuming that the states of the system can change
following a Markov process. Although MDP can model a broader range of
problems than bandit, it requires much more data to train and is
usually computationally expensive.

\subsection{Reinforcement learning in recommender systems}

Previous works have used reinforcement learning to recommend web
pages, travel information, books, and news etc. For example, Joachims
\textit{et al.} use Q-learning to guide users through web
pages~\cite{Joachims1997WebWatcher}.  Golovin \textit{et al.} propose
a general framework for web recommendation, as well as user implicit
feedback to update the
system~\cite{Golovin2004Reinforcement}. Zhang\textit{ et al. }propose
a personalized web-document recommender, where user profile is
represented as vector of terms. The weight of the terms are updated
based on the temporal difference method using both implicit and
explicit feedback~\cite{Zhang2001Personalized}.
In~\cite{Srivihok2005Ecommerce}, a Q-learning based travel recommender
is proposed, where trips are ranked using a linear function of several
attributes including trip duration, price and country, and the weights
are updated using user feedback. Shani \textit{et al.} use a MDP to
model the dynamics of user preference in book
recommendation~\cite{Shani2005}, where purchase history is used as the
states, and the generated profit is used as the payoffs.  Similarly,
in a web recommender~\cite{Taghipour2008Hybrid}, history web pages are
used as the states; web content similarity and user behavior are
combined as the payoffs.

In a seminal work done by~\cite{LiLihong10Contextual}, news are
represented as feature vectors; the click-through rates of news are
treated as the payoffs and assumed to be a linear function of news
feature vectors. A bandit model called LinUCB is proposed to learn the
weights of the linear function. Our work differs from this work in two
aspects. Fundamentally, music recommendation is different from news
recommendation due to the sequential relationship between
songs. Technically, the additional novelty factor of our rating model
makes the reward function nonlinear and the confidence bound difficult
to obtain. Therefore we need the Bayes-UCB approach and the
sophisticated Bayesian inference algorithms developed in
Section~\ref{sec:Bayesian-Regression-Models-And-Inference}.  Moreover,
we cannot apply the offline evaluation techniques developed
in~\cite{Li2011Unbiased} because we assume that ratings change
dynamically over time. As a result, we must conduct online evaluation
with real human subjects.

Although we believe reinforcement learning has great potential in
improving music recommendation, it has received relatively little
attention and found only limited application. Liu \textit{et al.}  use
MDP to recommend music based on a user's heart rate to help the user
maintain it within the normal range~\cite{Liu2009Music}. States are
defined as different levels of heart rate, and biofeedback is used as
payoffs. However, (1) parameters of the model are not learnt from
exploration, and thus exploration/exploitation tradeoff is not needed;
(2) the work does not disclose much information about the evaluation
of the approach. Chi \textit{et al.}  uses MDP to automatically
generate playlist~\cite{Chi2010Reinforcement}.  Both SARSA and
Q-learning are used to learn user preference, and, similar
to~\cite{Shani2005}, states are defined as mood categories of the
recent listening history.  However, in this work, (1)
exploration/exploitation tradeoff is not considered; (2) mood or
emotion, while useful, can only contribute so much to effective music
recommendation; and (3) the MDP model cannot handle long listening
history, as the state space grows exponentially with history length;
as a result, too much exploration and computation will be required to
learn the model. Independent of and concurrent with our work, Liebman
\textit{et al.} build a DJ agent to recommend playlists based on
reinforcement learning~\cite{DJ-MC}. Their work differs from ours in
that: (1) exploration/exploitation tradeoff is not considered; (2) the
reward function does not consider the novelty of recommendations; (3)
their approach is based on a simple tree-search heuristic, ours the
thoroughly studied muti-armed bandit; (4) not much information about
the simulation study is disclosed, and no user study is conducted.

The active learning approach developed by~\cite{Karimi2011Towards}
\textit{only} \textit{explores} songs in order to optimize the
predictive performance on a pre-determined test dataset. Our approach,
on the other hand, requires no test dataset and balances \textit{both}
exploration and exploitation to optimize the entire interactive
recommendation process between the system and users. Since many
recommender systems in reality do not have test data or at least have
no data for new users, our bandit approach is more realistic compared
with the active learning approach.

Our work is, to the best of our knowledge, the first to balance
exploration and exploitation based on reinforcement learning and
particularly multi-armed bandit in order to improve recommendation
performance and mitigate the cold-start problem in music
recommendation.

\section{A Bandit approach to music recommendation}

\label{sec:Modeling}

\subsection{Personalized user rating model }

\label{sub:The-rating-model}Music preference is a combined effect of
many factors including music audio content, novelty, diversity, moods
and genres of the songs, user emotional states, and user context
information~\cite{Wang2012Context}. As it is unrealistic to cover all
the factors in this paper, we focus on audio content and novelty.

\textbf{Music Audio Content} - Whether a user likes or dislikes a song
is highly related to its audio content. We assume that the music audio
content of a song can be described as a feature vector $\mbf{x}$.
Without considering other factors, a user's preference can be
represented as a linear function of $\mbf{x}$ as:
\begin{equation}
U_{c}=\bs{\theta}^{\prime}\mbf{x}\label{eq:factor_content}
\end{equation}
where the parameter vector $\bs{\theta}$ represents user
preference of different music features. Users may have different
preference and thus different values of $\bs{\theta}$. To keep
the problem simple, we assume a user's preference is invariant,
i.e. $\bs{\theta}$ remains a constant, and leave modeling
changing $\bs{\theta}$ as future work.

Although the idea of exploration/exploitation tradeoff can be applied
on collaborative filtering (CF) as long as the rating distribution can
be estimated as shown in
Figure~\ref{fig:Uncertainty-in-recommendation.}, we choose the
content-based approach instead of the popular CF-based methods for a
number of reasons. First, we need a posterior distribution of $U_{c}$
in order to use Bayes-UCB as introduced in
Section~\ref{sec:background}, so non-Bayesian methods cannot be
used. Second, existing Bayesian matrix factorization
methods~\cite{Salakhutdinov2008Bayesian,Silva2012Active} are much more
complicated than the linear model and also require large amount of
training data; these render the user study unwieldy and
expensive. Third, our bandit approach requires the model to be updated
whenever a new rating is obtained, but existing Bayesian matrix
factorization methods are too
slow~\cite{Salakhutdinov2008Bayesian,Silva2012Active}.  Fourth, CF
suffers from the new song problem while the content-based method does
not. Fifth, CF captures correlation instead of causality and thus does
not explain why a user likes a song. However, as science usually
pursues causal models, the content-based approach captures one
important aspect of the causality, i.e. music content.

\begin{figure}[t]
\centering
\begin{minipage}[b]{0.33\textwidth}
  \centering
  \includegraphics[bb=35bp 40bp 288bp 107bp,clip,scale=0.55]{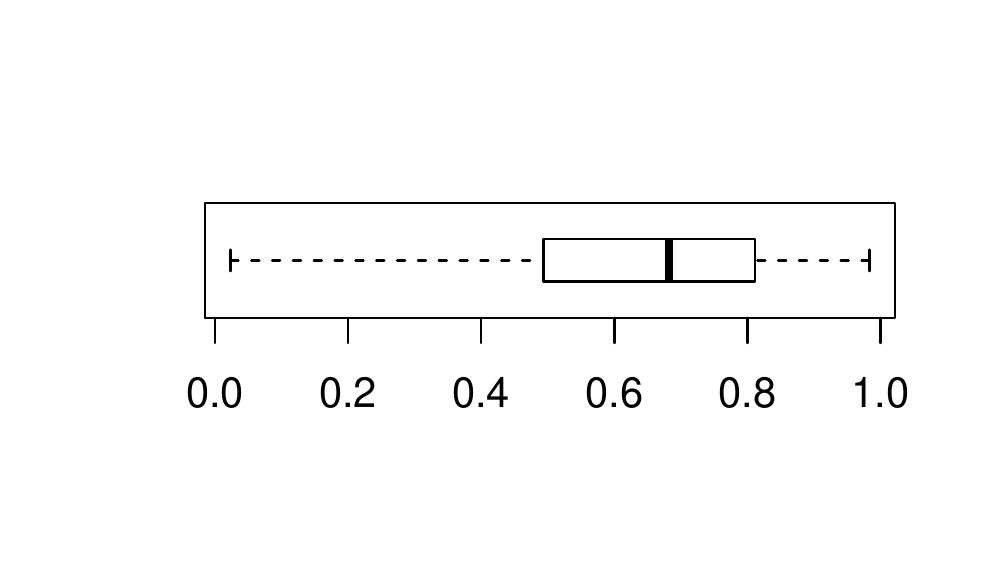}
  \vspace{4mm}
  \captionsetup{width=0.9\textwidth}
  \captionof{figure}{Proportion of repetitions in users' listen history}
  \label{fig:Proportion-of-repetitions}
\end{minipage}
\begin{minipage}[b]{0.33\textwidth}
  \centering
  \includegraphics[bb=15bp 10bp 340bp 208bp,clip,scale=0.44]{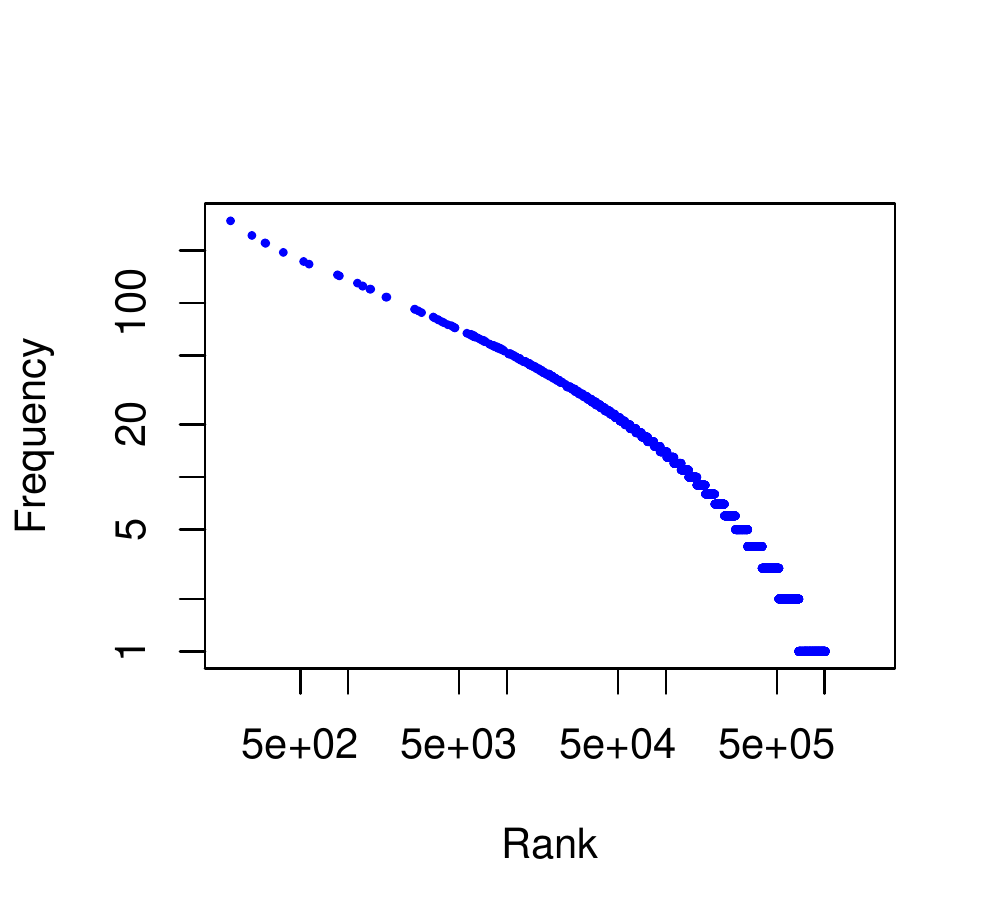}
  \captionsetup{width=0.9\linewidth}
  \captionof{figure}{Zipf's law of song repetition frequency}
  \label{fig:Repetition-frequency-of}
\end{minipage}
\begin{minipage}[b]{0.33\textwidth}
  \centering
  \includegraphics[bb=10bp 30bp 288bp 230bp,clip,scale=0.41]{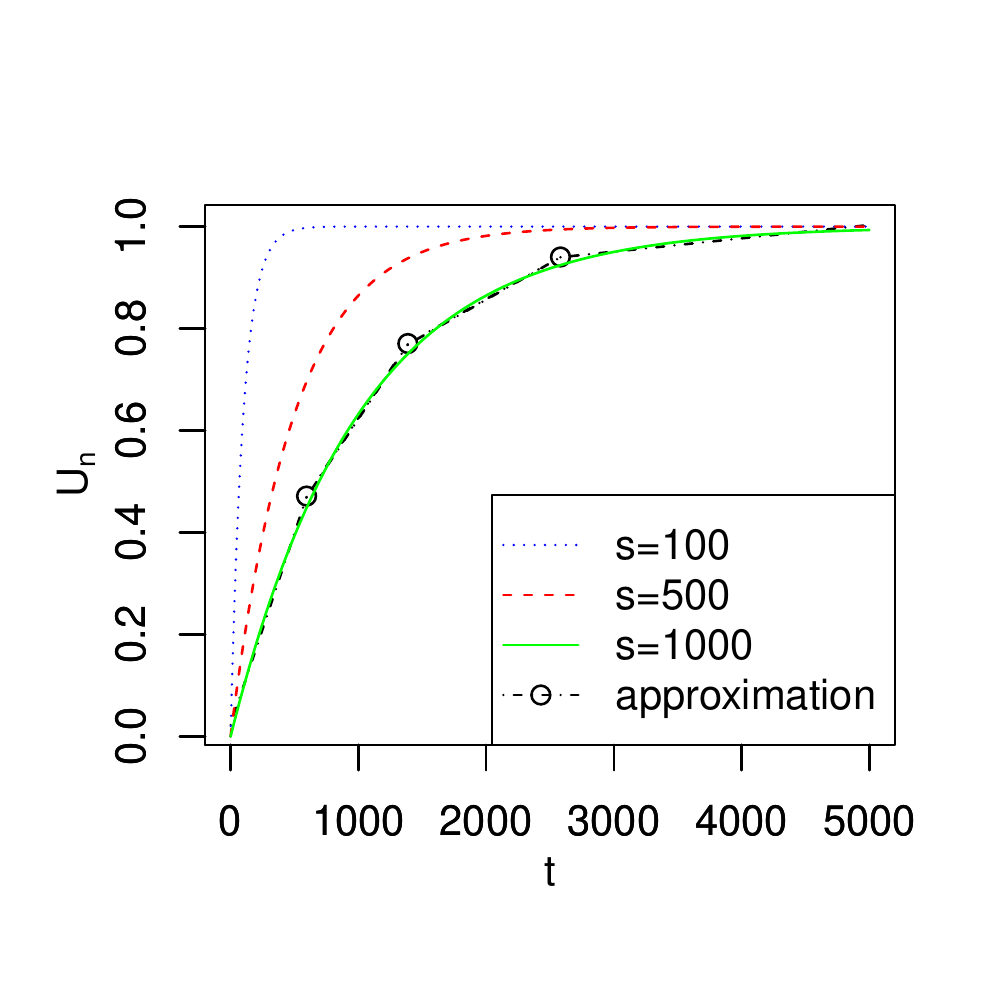}
  \captionsetup{width=0.95\linewidth}
  \vspace{-2mm}
  \captionof{figure}{Examples of $U_{n}=1-e^{-t/s}$. The line marked with circles is a
4-segment piecewise linear approximation.}
  \label{fig:u2u3}
\end{minipage}
\end{figure}

\textbf{Novelty }- We define that novelty is about repeating songs at
proper frequencies, which is in essence consistent with the definition
in~\cite{Gunawardana2009Survey}. We examined the repetition
distributions of 1000 users' listening histories collected from
Last.fm%
\footnote{http://ocelma.net/MusicRecommendationDataset/lastfm-1K.html%
}. The box plot in Figure~\ref{fig:Proportion-of-repetitions} shows
the proportion of repetitions, which is defined as: $
1-\frac{\mbox{number of of unique songs}}{\mbox{listening history
    length}} $.  Note that since Last.fm does not record users'
listening histories out side of Last.fm, the actual proportion should
be even larger than the 68.3\% shown here. Thus, most of the songs the
user listens to are repeats. We also studied the song repetition
frequency distribution of every individual user's listening history:
the frequencies of songs were first computed for every user; then all
users' frequencies were ranked in decreasing order; finally the
frequencies versus ranks were plotted on a log-log scale
(Figure~\ref{fig:Repetition-frequency-of}).  The distribution
approximately follows the Zipf's law~\cite{Newman2005Power}---only a
small set of songs are repeated for most of the time while all the
rest are repeated much less often. Most other types of recommenders,
however, do not follow Zipf's law. Recommending books that have been
bought or movies that have been watched makes little sense. In music
recommendation, however, it is critically important to repeat songs
appropriately.

Existing novelty models do not take time into
consideration~\cite{Lathia2010Temporal,Castells2011Novelty,Zhang2012Auralist},
and as a result songs heard year ago and just now have the same impact
on the current recommendation. Inspired by~\cite{Hu2011NEXTONE}, we
assume that the novelty of a particular song decays immediately after
listening to it and then gradually recovers. Let $t$ be the time
elapsed since the last listening of the song, the novelty recovers
following the function:
\begin{equation}
U_{n}=1-e^{-t/s}\label{eq:factor_novelty}
\end{equation}
where $s$ is a parameter indicating the recovery speed, with slower
recovery having a higher $s$. Figure~\ref{fig:u2u3} shows examples of
$U_{n}$ with different values of $s$.

Different users can have different recovery rates $s$. As can be seen
from the broad distribution in
Figure~\ref{fig:Proportion-of-repetitions}, some may repeatedly listen
to their favorite songs more often, while the others would be keen to
exploring new songs. Therefore we assume $s$ to be a personalized
value to be learnt through the user interactive process.

\textbf{Combined Model} - A user's preference of a recommendation can
be represented as a rating; the higher the rating is, the more the
user likes the recommendation. Unlike traditional recommenders which
assume ratings are static, we assume that a rating is the combined
effect of the user's preference of the song's content and the
dynamically changing novelty. Therefore, a song rated as 5 last time
could be rated as 2 this time because the novelty has
decreased. Finally, we define the complete user rating model as:
\begin{eqnarray}
  U & = & U_{c}U_{n}=\bs{\theta}^{\prime}\mbf{x}\left(1-e^{-t/s}\right).\label{eq:combined_rating}
\end{eqnarray}

In this model, the more the user likes a particularly song the more
likely it will be repeated---a song with larger $U_{C}$ requires less
time ($t$) to recover $U$ and becomes eligible for repeat.  Also,
given that the user's favorites comprise a small subset of his/her
library, the U model behaves in accordance with Zipf's Law and ensures
that only a small proportion of songs will be repeated often. This
property of the model will be verified in
Section~\ref{sub:Playlist-generation}.

In Section~\ref{sub:Recommendation-performance}, we will show that the
product form of Equation~(\ref{eq:combined_rating}) leads to
significantly better performance than the alternative linear
combination $U=aU_{c}+bU_{n}$.

\subsection{Interactive music recommendation}
\label{sub:Interactive-music-recommendation}

Under our rating model, each user is represented by a set of
parameters $\mbf{\Omega}=\{\mbf{\bs{\theta}},s\}$. If we
know the values of $\mbf{\Omega}$, we can simply recommend the
songs with the highest rating according to
Equation~(\ref{eq:combined_rating}).  However, $\mbf{\Omega}$ needs
to be estimated from historical data, and thus uncertainty always
exists. In this case, the greedy strategy used by traditional systems
is suboptimal, and it is necessary to take the uncertainty into
account and balance exploration and exploitation, as explained in
Section~\ref{sec:Introduction}.
\begin{figure}[t]
\begin{centering}
\includegraphics[bb=0bp 0bp 293bp 138bp,clip,scale=0.80]{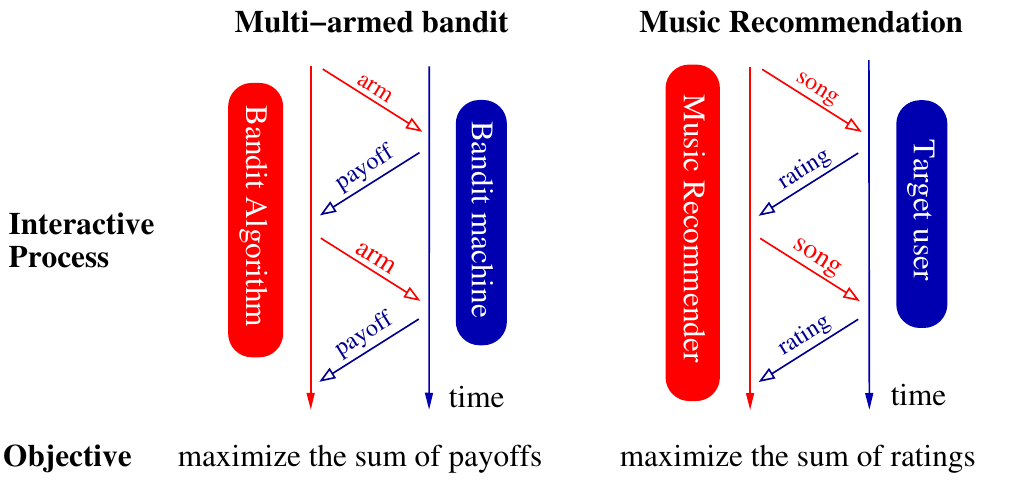}
\par\end{centering}

\caption{Relationship between the multi-armed bandit problem and music
  recommendation\label{fig:Relationship-between-multi-armed}}
\end{figure}

The multi-armed bandit approach introduced in
Section~\ref{sec:background} offers a way for balancing exploration
and exploitation for the interactive music recommendation process
between the target user and the recommender.  As illustrated in
Figure~\ref{fig:Relationship-between-multi-armed}, we treat songs as
arms, and user ratings as payoffs%
\footnote{Although in reality users usually do not give explicit
  feedback (i.e.  ratings) to every recommended song, implicit
  feedback (e.g. skipping a song, listening to a song fully) can be
  obtained much more easily.  In this paper, we focus on explicit
  feedback to keep the problem simple.%
}. The music recommendation problem is then transformed into a
multi-armed bandit problem, and the objective of a music recommender
is also changed to maximizing the sum of the ratings given by the
target user \emph{over the long term}. We argue that the cumulative
rating is a more realistic objective than the \emph{myopic} predictive
accuracy used by traditional music recomenders, because users usually
listen to songs for a long time instead of focusing on one individual
song.

We adopt the Bayes-UCB algorithm introduced in
Section~\ref{sec:background} for our recommendation task. First we
denote the rating given by the target user to recommendation $i$ as a
random variable $R_{i}$, and the expectation of $R_{i}$ is $U$ given
the feature vector $(\mbf{x}_{i},t_{i})$:
\begin{equation}
  \bb{E}[R_{i}]=U_{i}=\bs{\theta}^{\prime}\mbf{x}_{i}\left(1-e^{-t_{i}/s}\right)\label{eq:expected_rating}
\end{equation}

Then, we develop Bayesian models to estimate the posterior
distribution of $U$ given the history recommendation information. We
sketch the framework here and explain it in greater detail in
Section~\ref{sec:Bayesian-Regression-Models-And-Inference}.  We assume
that the prior distribution of $\mbf{\Omega}$ is
$p(\mbf{\Omega})$ and that, at the $(l+1)$-th recommendation, we
have accumulated $l$ history recommendations
$\mathcal{D}_{l}=\{(\mbf{x}_{i},t_{i},r_{i})\}_{i=1}^{l}$ as
training samples, where $r_{i}$ is the rating given by the user to the
$i$-th recommendation. The posterior distribution of $\mbf{\Omega}$
can then be obtained based on Bayes' rule:
\begin{align}
  p(\mbf{\Omega}|\mathcal{D}_{l})\ \propto\ &
  p(\mbf{\Omega})p(\mathcal{D}_{l}|\mbf{\Omega})\label{eq:parameters_posterior}
\end{align}
and then the expected rating of song $k$, denoted as $U_{k}$ can
be predicted as:
\begin{align}
  p(U_{k}|\mathcal{D}_{l})=\int
  p(U_{k}|\mbf{\Omega})p(\mbf{\Omega}|\mathcal{D}_{l})\mbf{d}\mbf{\Omega}\label{eq:posterior_expected_rating}
\end{align}
Later, we use $\lambda_{k}^{l}$ to denote $p(U_{k}|\mathcal{D}_{l})$
for simplicity.

Finally, to balance exploration and exploitation, Bayes-UCB recommends
song $k^{\ast}$, which maximizes the quantile function: 
\begin{align*} {\displaystyle
    k^{\ast}=\arg\max_{k=1\dots|\mathcal{S}|}Q(\alpha,\lambda_{k}^{l})}
\end{align*}
where $Q$ satisfies $\bb{P}\left[U_{k}\leq
  Q(\alpha,\lambda_{k}^{l})\right]=\alpha$ and $\mathcal{S}$ is all
songs in the database. We set $\alpha=1-\frac{1}{l+1}$.  The detail of
the recommendation algorithm is listed in
Algorithm~\ref{alg:Recommendation-using-Bayes-UCB}.

The cold-start problem is caused by the lack of information required
for making good recommendations. There are many ways for mitigating
the cold-start problem, most of which rely on additional information
about the users or songs, e.g., popularity/metadata information about
the songs~\cite{Hariri2012Contextaware}, context/demographic
information about the users~\cite{Wang2012Context}. Although music
audio content is required by $U_{c}$, it is usually easy to obtain in
industry.  Our bandit addresses the cold-start problem without relying
on additional information about users and songs. Instead, it wisely
explores and exploits information during the whole interactive
process. Thus, the bandit approach presents a fundamentally different
method to tackle the cold-start problem, yet it can be used in
conjunction with existing methods.

There are other Bayesian multi-arm bandit approaches such as Thompson
sampling~\cite{Agrawal2012Analysis} and optimistic Bayesian
sampling~\cite{May2012Optimistic}.  Theoretical performance
comparisons between them are interesting research problems. Empirical
results of them are, however, usually comparable.  These comparisons
are not the focus of this work. Moreover, since all of them are based
on the Bayesian approach, it is very easy to replace Bayes-UCB with
other approaches even if Bayes-UCB is shown to be inferior in the
future.

\begin{algorithm}
\caption{Recommendation using Bayes-UCB\label{alg:Recommendation-using-Bayes-UCB}}
\begin{algorithmic}
\FOR{$l=1$ to $n$}
	\FORALL{song $k=1,\dots,|\mathcal{S}|$}
        \STATE compute \quad $q_k^l=Q\left(1-1/l,  \lambda_k^{l-1}\right)$
	\ENDFOR
	\STATE recommend song $k^\ast = \arg\max_{k=1\dots |\mathcal{S}|}q_k^l$
	\STATE gather rating $r_l$; update $p(\mbf{\Omega}|\mathcal{D}_{l})$ and $\lambda_k^l$
\ENDFOR
\end{algorithmic}
\end{algorithm}

\section{Bayesian Models and Inference }

\label{sec:Bayesian-Regression-Models-And-Inference}

\subsection{Exact Bayesian model}

\label{sub:Bayesian-Regression-Models}
\begin{figure}[t]
\centering
\begin{minipage}{.48\textwidth}
\centering
\subfloat[Exact Bayesian model]{\makebox[\linewidth]{ \includegraphics[clip,scale=0.35]{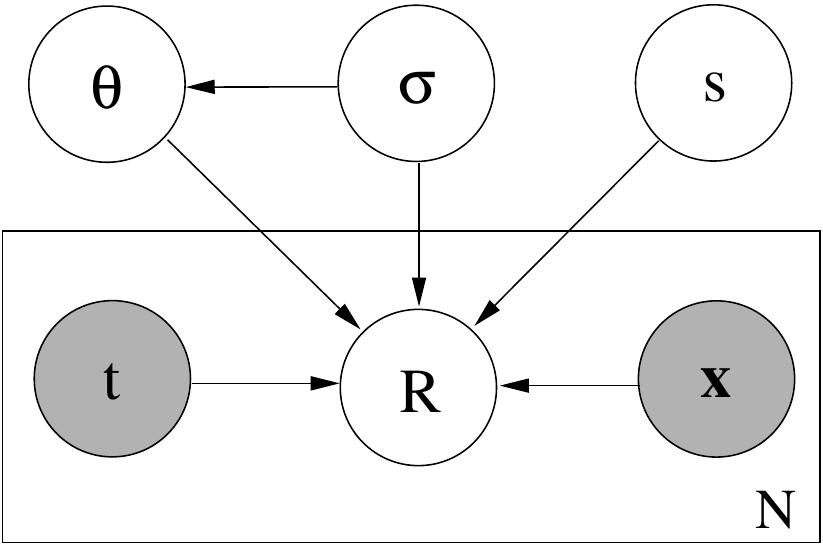}}
  \label{fig:Bayesian-Model}}
\end{minipage} 
\begin{minipage}{.48\textwidth}
\centering
\subfloat[Approximate Bayesian model]{\makebox[\linewidth]{ \includegraphics[scale=0.35]{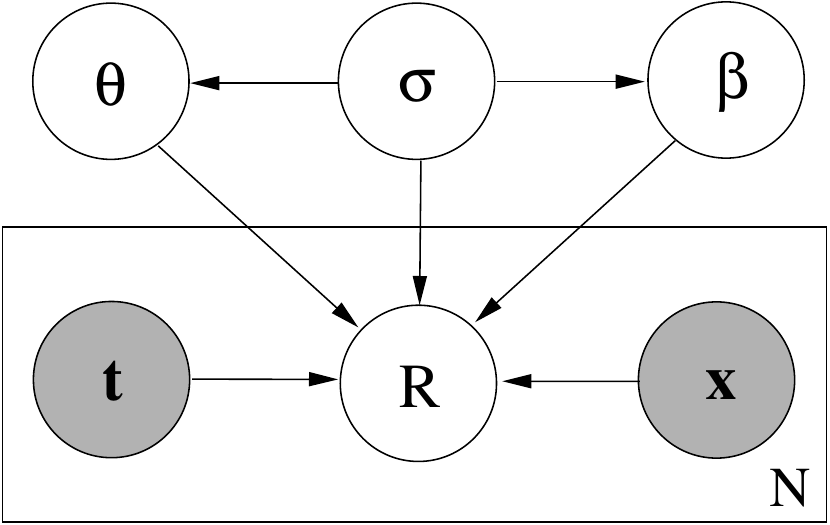}}
  \label{fig:Approximate-Bayesian-Model}}
\end{minipage} 

\caption{Graphical representation of the Bayesian models. Shaded nodes
  represent observable random variables, while white nodes represent
  hidden ones.  The rectangle (plate) indicates that the nodes and
  arcs inside are replicated for $N$ times.}
\end{figure}

To compute
Equations~(\ref{eq:parameters_posterior})
and~(\ref{eq:posterior_expected_rating}), we develop the following
Bayesian model with its graphical representation shown in
Figure~\ref{fig:Bayesian-Model}.
\begin{align*}
  R|\mbf{x},t,\bs{\theta},s,\sigma^{2} & \sim\mathcal{N}\left(\bs{\theta}^{\prime}\mbf{x}\left(1-e^{-t/s}\right),\sigma^{2}\right)\\
  \bs{\theta}|\sigma^{2} & \sim\mathcal{N}(\bs{0},a_{0}\sigma^{2}\mbf{I})\\
  s & \sim\mathcal{G}(b_{0,}c_{0})\\
  \tau=1/\sigma^{2} & \sim\mathcal{G}(f_{0},h_{0})
\end{align*}

Every line of the model indicates a probability dependency and the
corresponding distribution, e.g.,
$\bs{\theta}|\sigma^{2}\sim\mathcal{N}(\bs{0},a_{0}\sigma^{2}\mbf{I})$
suggests
$p(\bs{\theta}|\sigma^{2})=\mathcal{N}(\bs{0},a_{0}\sigma^{2}\mbf{I})$.
$\mathcal{N}(\cdot,\cdot$) is a (multivariate) Gaussian distribution
with the mean and (co)variance parameters, and
$\mathcal{G}(\cdot,\cdot)$ is a Gamma distribution with the shape and
rate parameters. The rating $R$ is assumed to be normally distributed
following the convention of recommender systems. A gamma prior is put
on $s$ because $s$ is positive. Following the conventions of Bayesian
regression models, A normal prior is put on $\bs{\theta}$ and
a gamma one for $\tau$. We depend on $\sigma^{2}$ for
$\bs{\theta}$ because it shows better convergence in the
simulation study.

Since there is no closed form solution to
Equation~(\ref{eq:parameters_posterior}) for this model, Markov Chain
Monte Carlo (MCMC) is used as the approximate inference
procedure. Directly evaluating
Equation~(\ref{eq:posterior_expected_rating}) is also impossible. Thus
we use Monte Carlo simulation to obtain $\lambda_{k}^{l}$: for every
sample obtained from the MCMC procedure, we substitute it into
Equation~(\ref{eq:expected_rating}) to obtain a sample of $U_{i}$, and
then use the histogram of the samples of $U_{i}$ as an approximation
of $\lambda_{k}^{l}$.

This approach is easy to understand and implement. However, it is very
slow and users can hardly wait for a recommendation for tens of
seconds until the Markov chain converges. To make the algorithm more
responsive, we will develop an approximate Bayesian model and a highly
efficient variational inference algorithm in the following sections.

\subsection{Approximate Bayesian model}

\label{sec:approximate_model}

\subsubsection{Piecewise Linear approximation }

\label{sub:Continuous-Piecewise-Linear-approximation}It is very
difficult to develop better inference algorithms for the exact
Bayesian model because of the irregular form of function
$U_{n}(t)$. Fortunately, we find $U_{n}$ can be approximated by a
piecewise linear function (as shown in Figure~\ref{fig:u2u3}), which
enables us to develop an efficient model.

For simplicity, we discretize time $t$ into $K$ predetermined
intervals: $[0,\xi_{1}),[\xi_{1},\xi_{2}),\dots[\xi_{K-1},+\infty)$,
and only consider the class of piecewise linear functions whose
consecutive line segments intersect at the boundaries of the
intervals. It is not difficult to see that this class of functions can
be compactly represented as a linear
function~\cite{hastie2009elements}. We first map $t$ into a vector
$\bs{t}=[(t-\xi_{1})_{+},\dots(t-\xi_{K-1}),t,1]$, where
$(t-\xi)_{+}=\max(t-\xi,0)$, and then approximate $U_{n}(t)$ as $
U_{n}(t)\approx\bs{\beta}^{\prime}\bs{t} $, where
$\bs{\beta}=[\beta_{1},\dots\beta_{K+1}]^{\prime}$ is a vector of
parameters to be learnt from training data. Now, we can represent $U$
as the product of two linear functions: $ U
=U_{c}U_{n}\approx\bs{\theta}^{\prime}\mbf{x}\bs{\beta}^{\prime}\bs{t}$.

Based on this approximation, we revise the distributions of $R$ and
the parameters of the exact Bayesian model as follows:
\begin{align}
  R|\mbf{x},\mbf{t},\bs{\theta},\bs{\beta},\sigma^{2} & \sim\mathcal{N}(\bs{\theta}^{\prime}\mbf{x}\bs{\beta}^{\prime}\mbf{t},\sigma^{2})\label{eq:R_x_t_theta_beta_sigma}\\
  \bs{\theta}|\sigma^{2} & \sim\mathcal{N}(\bs{\mu}_{\bs{\theta}0},\sigma^{2}\mbf{D}_{0})\nonumber \\
  \bs{\beta}|\sigma^{2} & \sim\mathcal{N}(\bs{\mu}_{\bs{\beta}0},\sigma^{2}\mbf{E}_{0})\nonumber \\
  \tau=1/\sigma^{2} & \sim\mathcal{G}(a_{0},b_{0})\nonumber
\end{align}
where $\bs{\theta},\bs{\beta},\tau$ are parameters.
$\mbf{D}_{0},\mbf{E}_{0},\bs{\mu}_{\bs{\theta}0},\bs{\mu}_{\bs{\beta}0},a_{0},b_{0}$
are hyperparameters of the priors to be specified beforehand. $\mbf{D}_{0}$
and $\mbf{E}_{0}$ are positive definite matrices. The graphical
representation of the model is shown in Figure~\ref{fig:Approximate-Bayesian-Model}.
We use conjugate priors for $\bs{\theta},\bs{\beta},\tau$,
which make the variational inference algorithm described later very
efficient.

\subsubsection{Variational inference }

\label{sub:Variational-Inference} Recall that our objective is to
compute the posterior distribution of parameters $\mbf{\Omega}$
(now it is $\{\bs{\theta},\bs{\beta},\tau\}$) given
the history data
$\mathcal{D}=\{(\mbf{x}_{i},\mbf{t}_{i},r_{i})\}_{i=1}^{N}$,
i.e., $p(\bs{\theta},\bs{\beta},\tau|\mathcal{D})$.
Using piecewise linear approximation, we can now develop an efficient
variational inference algorithm.

Following the convention of mean-field
approximation~\cite{Friedman2009Probabilistic}, we assume that the
joint posterior distribution can be approximated by a restricted
distribution $q(\bs{\theta},\bs{\beta},\tau)$, which
consists of three independent
factors~\cite{Friedman2009Probabilistic}:
\[
p(\mbf{\Omega}|\mathcal{D})=p(\bs{\theta},\bs{\beta},\tau|\mathcal{D})\approx
q(\bs{\theta},\bs{\beta},\tau)=q(\bs{\theta})q(\bs{\beta})q(\tau).
\]
Because of the choice of the conjugate priors, it is easy to show that
the restricted distributions $q(\bs{\theta})$,
$q(\bs{\beta})$, and $q(\tau)$ take the same parametric forms
as the prior distributions.  Specifically,
\begin{eqnarray*}
  q(\bs{\theta}) \propto\exp\left(-\frac{1}{2}\bs{\theta}^{\prime}\bs{\Lambda}_{\bs{\theta}N}\bs{\theta}+\bs{\eta}_{\bs{\theta}N}^{\prime}\bs{\theta}\right), & 
  q(\bs{\beta}) \propto\exp\left(-\frac{1}{2}\bs{\beta}^{\prime}\bs{\Lambda}_{\bs{\beta}N}\bs{\beta}+\bs{\eta}_{\bs{\beta}N}^{\prime}\bs{\beta}\right), &
  q(\tau) \propto\tau^{a_{N}-1}\exp\left(-b_{N}\tau\right).
\end{eqnarray*}

To find the values that minimize the KL-divergence between
$q(\bs{\theta},\bs{\beta},\tau)$ and the true
posterior $p(\bs{\theta},\bs{\beta},\tau|\mathcal{D})$
for parameters $\bs{\Lambda}_{\bs{\theta}N}$,
$\bs{\eta}_{\bs{\theta}N}$,
$\bs{\Lambda}_{\bs{\beta}N}$,
$\bs{\eta}_{\bs{\beta}N}$, $a_{N}$, and $b_{N}$, we
use the coordinate descent method. Specifically, we first initialize
the parameters of $q(\bs{\theta})$,
\textbf{$q(\bs{\beta})$}, and \textbf{$q(\tau)$}, and then
iteratively update $q(\bs{\theta})$,
\textbf{$q(\bs{\beta})$}, and \textbf{$q(\tau)$} until the
variational lower bound $\mathcal{L}$ (elaborated in the
Appendix
) converges. Further explanation about the principle can be found
in~\cite{Friedman2009Probabilistic}. The detailed steps are in
Algorithm~(\ref{alg:variational-inference}), where $p$, $K$ are the
dimensionalities of $\mbf{x}$ and $\mbf{t}$, respectively; the
moments of $\bs{\theta},\bs{\beta},\tau$ are in the
Appendix.

\newcommand*{\Scale}[2][4]{\scalebox{#1}{$#2$}}%
\begin{algorithm}
\caption{Variational inference\label{alg:variational-inference}}
\begin{algorithmic}
\STATE \textbf{input:} 
                $\mathcal{D},           
                \mbf{D}_{0},
                \mbf{E}_{0},
                \bs{\mu}_{\bs{\theta}0},
                \bs{\mu}_{\bs{\beta}0},a_{0},b_{0}$
\STATE initialize 
                $\bs{\Lambda}_{\bs{\theta}N}$, 
                $\bs{\eta}_{\bs{\theta}N}$, 
                $\bs{\Lambda}_{\bs{\beta}N}$, 
                $\bs{\eta}_{\bs{\beta}N}$, $a_{N}$, $b_{N}$
\REPEAT
        \STATE update $q(\bs{\theta})$: \quad $ \bs{\Lambda}_{\bs{\theta}N}  \leftarrow\bb{E}[\tau]\left(\mbf{D}_{0}^{-1}+\sum_{i=1}^{N}\mbf{x}_{i}\mbf{t}_{i}^{\prime}\bb{E}\left[\bs{\beta}\bs{\beta}^{\prime}\right]\mbf{t}_{i}\mbf{x}_{i}^{\prime}\right),
\quad \bs{\eta}_{\bs{\theta}N} \leftarrow\bb{E}[\tau]\left(\mbf{D}_{0}^{-1}\bs{\mu}_{\bs{\theta}0}+\sum_{i=1}^{N}r_{i}\mbf{x}_{i}\mbf{t}_{i}^{\prime}\bb{E}[\bs{\beta}]\right) $

        \STATE update $q(\bs{\beta})$: \quad $ \bs{\Lambda}_{\bs{\beta}N}  \leftarrow\bb{E}[\tau]\left(\mbf{E}_{0}^{-1}+\sum_{i=1}^{N}\mbf{t}_{i}\mbf{x}_{i}^{\prime}\bb{E}[\bs{\theta}\bs{\theta}^{\prime}]\mbf{x}_{i}\mbf{t}_{i}^{\prime}\right),
 \quad  \bs{\eta}_{\bs{\beta}N} \leftarrow\bb{E}[\tau]\left(\mbf{E}_{0}^{-1}\bs{\mu}_{\bs{\beta}0}+\sum_{i=1}^{N}r_{i}\mbf{t}_{i}\mbf{x}_{i}^{\prime}\bb{E}[\bs{\theta}]\right) $

        \STATE update $q(\tau)$: \quad $ a_{N} \leftarrow\frac{p+K+N}{2}+a_{0} $,
\begin{align*} b_{N} & \leftarrow\frac{1}{2}\left[\tr\left[D_{0}^{-1}\left(\bb{E}[\bs{\theta}\bs{\theta}^{\prime}]\right)\right]+\left(\bs{\mu}_{\bs{\theta}0}^{\prime}-2\bb{E}[\bs{\theta}]^{\prime}\right)\mbf{D}_{0}^{-1}\bs{\mu}_{\bs{\theta}0}\right]
 +\frac{1}{2}\left[\tr\left[E_{0}^{-1}\left(\bb{E}[\bs{\beta}\bs{\beta}^{\prime}]\right)\right]+\left(\bs{\mu}_{\bs{\beta}0}^{\prime}-2\bb{E}[\bs{\beta}]^{\prime}\right)\mbf{E}_{0}^{-1}\bs{\mu}_{\bs{\beta}0}\right]\\  
& +\frac{1}{2}\sum_{i=1}^{N}\left(r_{i}^{2}+\mbf{x}_{i}^{\prime}\bb{E}[\bs{\theta}\bs{\theta}^{T}]\mbf{x}_{i}\mbf{t}_{i}^{\prime}\bb{E}[\bs{\beta}\bs{\beta}^{T}]\mbf{t}_{i}\right)  
 -\sum_{i=1}^{N}r_{i}\mbf{x}_{i}^{\prime}\bb{E}[\bs{\theta}]\mbf{t}_{i}^{\prime}\bb{E}[\bs{\beta}]+b_{0} \end{align*} 

\UNTIL{$\mathcal{L}$ converges}
\RETURN 
                $\bs{\Lambda}_{\bs{\theta}N}$, 
                $\bs{\eta}_{\bs{\theta}N}$, 
                $\bs{\Lambda}_{\bs{\beta}N}$, 
                $\bs{\eta}_{\bs{\beta}N}$, $a_{N}$, $b_{N}$
\end{algorithmic}
\end{algorithm}

\subsubsection{Predict the posterior distribution $p(U|\mathcal{D})$}

Because $q(\bs{\theta})$ and $q(\bs{\beta})$ are
normal distributions, $\bs{\theta}^{\prime}\mbf{x}$ and
$\bs{\beta}^{\prime}\mbf{t}$ are also normally distributed:
\begin{align*}
  p(\bs{\theta}^{\prime}\mbf{x}|\mbf{x},\mbf{t},\mathcal{D})
  \approx\mathcal{N}(\mbf{x}^{\prime}\bs{\Lambda}_{\bs{\theta}N}^{-1}\bs{\eta}_{\bs{\theta}N},\mbf{x}^{\prime}\bs{\Lambda}_{\bs{\theta}N}^{-1}\mbf{x}),
  & \quad &
  p(\bs{\beta}^{\prime}\mbf{t}|\mbf{x},\mbf{t},\mathcal{D})
  \approx\mathcal{N}(\mbf{t}^{\prime}\bs{\Lambda}_{\bs{\beta}N}^{-1}\bs{\eta}_{\bs{\beta}N},\mbf{t}^{\prime}\bs{\Lambda}_{\bs{\beta}N}^{-1}\mbf{t})
\end{align*}
and the posterior distribution of $U$ in
Equation~(\ref{eq:posterior_expected_rating}) can be computed as:
\begin{align*}
  p(U|\mbf{x},\mbf{t},\mathcal{D}) &
  =p(\bs{\theta}^{\prime}\mbf{x}\bs{\beta}^{\prime}\mbf{t}|\mbf{x},\mbf{t},\mathcal{D})
  =\int
  p(\bs{\theta}^{\prime}\mbf{x}=a|\mbf{x},\mbf{t},\mathcal{D})p(\bs{\beta}^{\prime}\mbf{t}=\frac{U}{a}|\mbf{x},\mbf{t},\mathcal{D})da.
\end{align*}
Since there is no closed-form solution to the above integration, we
use Monte Carlo simulation: we first obtain one set of samples for
each of $\bs{\theta}^{\prime}\mbf{x}$ and
$\bs{\beta}^{\prime}\mbf{t}$, and then use the element-wise
products of the two group of samples to approximate the distribution
of $U$. Because $\bs{\theta}^{\prime}\mbf{x}$ and
$\bs{\beta}^{\prime}\mbf{t}$ are normally distributed
univariate random variables, the sampling can be done very
efficiently.  Moreover, prediction for different songs is trivially
parallelizable and is thus scalable.

\subsubsection{Integration of other factors}
\label{sub:Integration-of-other-utility-functions}

Although the approximate model considers music audio content and
novelty only, it is easy to integrate other factors as long as they
can be approximated by linear functions. For instance, diversity is
another important factor for a playlist. If we measure the diversity
that a song contributes to a playlist as $d$, and user preference of
$d$ follows a function that can be approximated by a piecewise linear
function. Following the method in
Section~\ref{sub:Continuous-Piecewise-Linear-approximation}, we can
map $d$ into a vector $\mbf{d}$ and modify the approximate Bayesian
model in Section~(\ref{sub:Continuous-Piecewise-Linear-approximation})
by extending Equation~(\ref{eq:R_x_t_theta_beta_sigma}) with an
additional term $\bs{\gamma}^{\prime}\mbf{d}$ and put a
prior on $\bs{\gamma}$ as following:
\begin{align*}
  R|\mbf{x},\mbf{t},\mbf{d},\sigma^{2},\bs{\theta},\bs{\beta},\gamma \sim\mathcal{N}(\bs{\theta}^{\prime}\mbf{x}\bs{\beta}^{\prime}\mbf{t}\bs{\gamma}^{\prime}\mbf{d},\sigma^{2}) &, \quad &
  \bs{\gamma}|\sigma^{2} \sim\mathcal{N}(\bs{\mu}_{\bs{\gamma}0},\sigma^{2}\mbf{F}_{0}).
\end{align*}
Following the symmetry between $\mbf{x}$, $\mbf{t}$, and $\mbf{d}$,
we can easily modify Algorithm~\ref{alg:variational-inference} accordingly
without further derivation. 

Similarly, we could incorporate in the model more factors such as
coherence of mood and genre. Moreover, although the model is designed
for music recommendation, it can also be applied for other regression
as long as the regression function can be factorized into the product
of a few linear functions.

\section{Experiments }

\label{sec:Experiments} We compare the results from our evaluations of
6 recommendation algorithms in this section. Extensive experimental
evaluations of both efficiency and effectiveness of the algorithms and
models have been conducted, and the results show significant promise
from both aspects.

\subsection{Experiment setup}

\subsubsection{Comparison recommendation algorithms}

To study the effectiveness of the exploration/exploitation tradeoff,
we introduced the Random and Greedy baselines. The Random approach
represents pure exploration and recommends songs uniformly at random.
The Greedy approach represents pure exploitation and always recommends
the song with the highest predicted rating. Therefore, the Greedy
approach simulates the strategy used by the traditional recommenders.
For Greedy, minimum mean square error approach was used to estimate
the parameters $\{\bs{\theta},s\}$, which were optimized by
the L-BFGS-B algorithm~\cite{Byrd1995Limited}.

To study the effectiveness of the rating model, the LinUCB baseline
was introduced. LinUCB is a bandit algorithm which assumes that the
expected rating is a linear function of the feature
vector~\cite{LiLihong10Contextual}.  In LinUCB, ridge regression is
used as the regression method, and upper confidence bound is used to
balance exploration and exploitation.

Two combinations of the factors $U_{c}$, $U_{n}$ were evaluated:
$U_{c}$ and $U_{c}U_{n}$. We write them as C and CN for short, where C
and N indicate content and novelty respectively, e.g., Bayes-UCB-CN
contains both content and novelty.

For the Bayes-UCB algorithm, the exact Bayesian model with the MCMC
inference algorithm (Section~\ref{sub:Bayesian-Regression-Models}) is
indicated by Bayes-UCB-CN, and the approximate model with the
variational inference algorithm (Section~\ref{sec:approximate_model})
is indicated by Bayes-UCB-CN-V.

We evaluated 6 recommendation algorithms, which were combinations of
the four approaches and three factors: Random, LinUCB-C, LinUCB-CN,
Bayes-UCB-CN, Bayes-UCB-CN-V, and Greedy-CN. Because LinUCB-CN cannot
handle nonlinearity and thus cannot directly model $U_{c}U_{n}$, we
combined the feature vector $\mbf{x}$ in $U_{c}$ and the time variable
$t$ in $U_{n}$ as one vector, and assumed the expected rating is a
linear function of the combined vector. Greedy-C was not included
because it was not related to our objective. As discussed in
Section~\ref{sub:Interactive-music-recommendation}, the bandit
approach can also combine with existing methods to solve the
cold-start problem. We plan to study the effectiveness of such
combinations in future works.

\subsubsection{Songs and Features}

Ten thousand songs from different genres were used in the experiments.
Videos of the songs were first crawled from YouTube and converted by
ffmpeg%
\footnote{http://ffmpeg.org%
} into mono channel WAV files with a 16KHz sampling rate. For every
song, a 30-second audio clip was used~\cite{Wang2012Context}. Feature
vectors were then extracted using a program developed based on the
MARSYAS library%
\footnote{http://marsyas.sourceforge.net%
}, in which a window size of 512 was used without overlapping. The
features we used and their dimensionalities are ZeroCrossing (1),
Centroid (1), Rolloff (1), Flux (1), MFCC (13), Chroma (14), SCF (24)
and SFM (24). These features are well accepted in the music
retrieval/recommendation domain. To represent a 30-second clip in one
feature vector, we used the mean and standard deviation of all feature
vectors from the clip.  Next, we added the 1-dimensional feature
\emph{tempo} to the summarized feature vectors, and the resulting
feature dimensionality is $79\times2+1=159$. Directly using the
$159$-dimensional features requires a large amount of data to train
the models and makes user studies very expensive and
time-consuming. To reduce the dimensionality, we conducted Principal
Component Analysis (PCA) with $90\%$ of variance reserved. The final
feature dimensionality is thus reduced to $91$.

The performance of these features in music recommendation was checked
based on a dataset that we built. We did not use existing music
recommendation datasets because they lack explicit ratings, and
dealing with implicit feedbacks is not our focus. Fifty-two
undergraduate students with different cultural backgrounds contributed
to the dataset, with each student annotating 400 songs with a 5-point
Likert scale from {}``very bad'' (1) to {}``very good'' (5). We first
computed the 10-fold cross-validation RMSE of $U_{c}$ for each
user. We then averaged the accuracy over all users. The resulting RMSE
is $1.10$, significantly lower than the RMSE ($1.61$) of the random
baseline with the same distribution as the data. Therefore these audio
features indeed provide useful information for recommendation. Feature
engineering can improve the accuracy, but it is not our focus and we
leave it as future work.

\subsubsection{Evaluation protocol}

\label{sub:Most-existing-recommenders}In~\cite{Li2011Unbiased}, an
offline approach is proposed for evaluating contextual-bandit
approaches with the assumption that the context (including the audio
features and the elapsed time of songs) at different iterations are
identically independently distributed. Unfortunately, this is not true
in our case because when a song is not recommended, its elapsed time
$t$ keeps increasing and is thus strongly correlated. Therefore,
online user study is the most reliable way of evaluation.

To reduce the cost of the user study, we first conducted comprehensive
simulation study to verify the approaches. Only if they passed the
simulations, we then proceeded to user study for further verification.
The whole process underwent for a few iterations, during which the
models and algorithms were continually refined. The results hereby
presented are from the final iteration, and intermediate results are
either referred to as preliminary study whenever necessary or omitted
due to page limitation.

\subsection{Simulations}

\subsubsection{Effectiveness study }

\label{sub:Effectiveness-study} 

$U=U_{c}U_{n}$ was used as the true model because the preliminary user
studies showed that this resulted in better performance, which will be
verified in Section~\ref{sub:User-study} again. During the simulation,
songs were recommended and rated about every $50$ seconds.  After 20
songs, the simulation paused for about 4 minutes to simulate the gap
between two recommendation sessions.

Priors for the Bayesian models were set as uninformative ones or
chosen based on preliminary simulation and user studies. For the exact
Bayesian model, they are: $a_{0}=10$, $b_{0}=3$, $c_{0}=10^{-2}$,
$f_{0}=10^{-3}$, $h_{0}=10^{-3}$, where $f_{0},h_{0}$ are
uninformative and $a_{0},b_{0},c_{0}$ are based on preliminary
studies. For the approximate Bayesian model, they are:
$\mbf{D}_{0}=\mbf{E}_{0}=10^{-2}\mbf{I}$,
$\bs{\mu}_{\bs{\theta}0}=\bs{\mu}_{\bs{\beta}0}=\mbf{0}$,
$a_{0}=2,b_{0}=2\times10^{-8}$, where
$\bs{\mu}_{\bs{\theta}0},\bs{\mu}_{\bs{\beta}0},a_{0},b_{0}$ are
uninformative and $\mbf{D}_{0},\mbf{E}_{0}$ are based on preliminary
studies; $\mbf{I}$ is the identity matrix.

$U_{n}$ was discretized into the following intervals (in minutes)
according to the exponentially decaying characteristics of human memory~\cite{ebbinghaus1913memory}:
$ [0,2^{-3}),[2^{-3},2^{-2}),\dots,[2^{10},2^{11}),[2^{11},+\infty)$.
We defined the smallest interval as $[0,2^{-3})$ because people
usually don't listen to a song for less than $2^{-3}$ minute. The
largest interval was defined as $[2^{11},+\infty)$ because our
preliminary user study showed that evaluating one algorithm takes no
more than $1.4$ day, i.e., about $2^{11}$ minutes. Further
discretization of $[2^{11},+\infty)$ should be easy. For songs that
had not been listened to by the target user, the elapsed time $t$ was
set as one month to ensure the $U_{n}$ is close to $1$.

We compared the recommendation performance of the 6 recommendation
algorithms in terms of regret, which is a widely used metric in RL
literatures. First we define that for the $l$-th recommendation, the
difference between the maximum expected rating
$ \bb{E}\left[\hat{R}^{l}\right]=\max_{k=1\dots|\mathcal{S}|}U_{k} $
and the expected rating of the recommended song is
$\Delta_{l}=\bb{E}\left[\hat{R}^{l}\right]-\bb{E}\left[R^{l}\right]$.
Then, the regret till the $n$-th recommendation can be written as
Equation~\ref{eq:regret}, where a smaller $\mathfrak{R}_{n}$ indicates
better performance.
\begin{align}
\mathfrak{R}_{n} & =\sum_{l=1\dots n}\Delta_{l}=\sum_{l=1\dots n}\bb{E}\left[\hat{R}^{l}\right]-\bb{E}\left[R^{l}\right]\label{eq:regret}
\end{align}

\begin{figure}
\centering
\begin{minipage}{0.48\textwidth}
\centering
\includegraphics[bb=0bp 0bp 396bp 346bp,clip,scale=0.45]{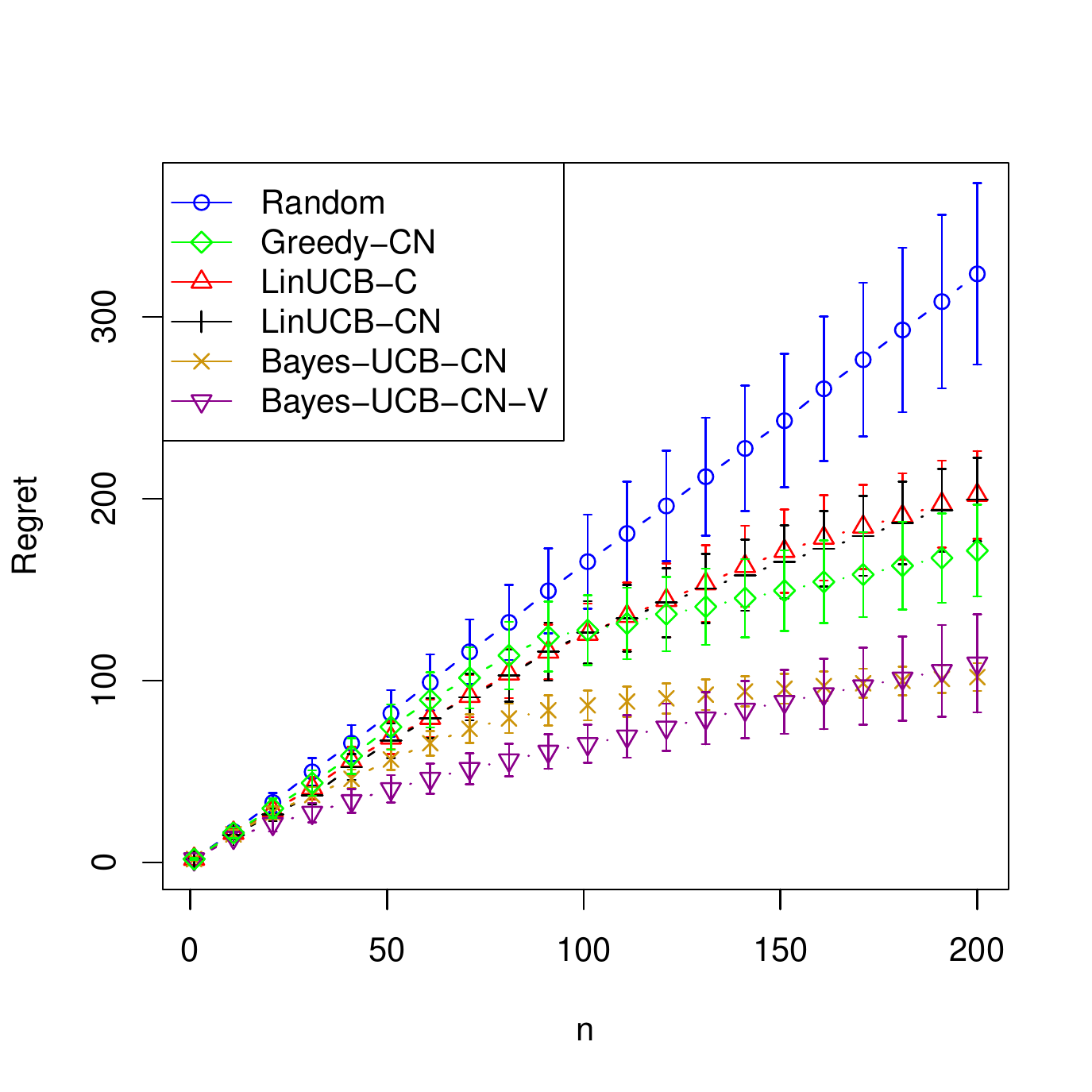}
\vspace{-5mm}
\captionof{figure}{Performance comparison in simulation}
\label{fig:Effectiveness-comparison}
\end{minipage}
\begin{minipage}{0.48\textwidth}
\centering
\includegraphics[bb=0bp 0bp 396bp 346bp,clip,scale=0.45]{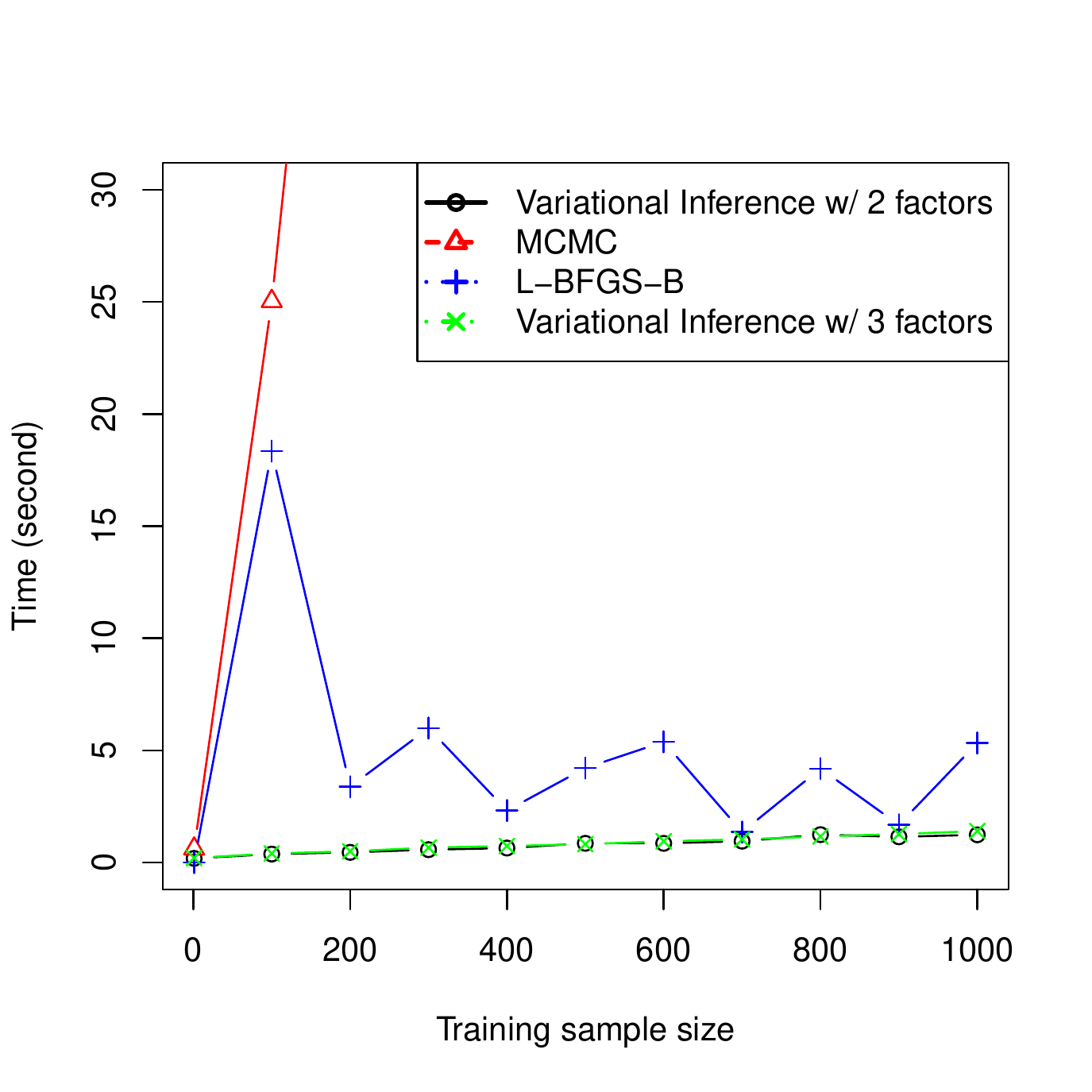}
\vspace{-5mm}
\captionof{figure}{Efficiency comparison}
\label{fig:Efficiency-comparison-between}
\end{minipage}
\end{figure}

Different values of parameters $\{\mbf{\bs{\theta}},s\}$
were tested. Elements of $\mbf{\bs{\theta}}$ were sampled
from standard normal distribution and $s$ was sampled from
$\mbox{uniform}(100,1000)$, where the range $(100,1000)$ was
determined based on preliminary user study. We conducted $10$ runs of
the simulation study. Figure~\ref{fig:Effectiveness-comparison} shows
the means and standard errors of the regret of different algorithms at
different number of recommendations $n$. From the figure, we see that
the algorithm Random (pure exploration) performs the worst. The two
LinUCB-based algorithms are worse than Greedy-CN because LinUCB-C does
not capture the novelty and LinUCB-CN does not capture the
nonlinearity within $U_{c}$ and $U_{n}$ although both LinUCB-C and
LinUCB-CN balance exploration and exploitation.

Bayes-UCB-based algorithms performed better than Greedy-CN because
Bayes-UCB balances exploration and exploitation. In addition, the
difference between Bayes-UCB and Greedy increases very fast when $n$
is small. This is because small $n$ means small number of training
samples and thus high uncertainty, i.e., the cold-start stage. Greedy
algorithms, which are used by most existing recommendation systems, do
not handle the uncertainty well, while Bayes-UCB can reduce the
uncertainty quickly and thus improves the recommendation performance.
The good performance of Bayes-UCB-CN-V also indicates that the
piecewise linear approximation and variational inference is accurate.

\subsubsection{Efficiency study}

Theoretical efficiency study of MCMC and variational inference
algorithms are difficult to analyze due to their iterative nature and
deserve future work. Instead, we conducted empirical efficiency study
of the training algorithms for Bayes-UCB-CN (MCMC), Bayes-UCB-CN-V
(variational inference), Greedy-CN (L-BFGS-B). In addition, the
variational inference algorithm for the 3-factor model describe in
Section~\ref{sub:Integration-of-other-utility-functions} was also
studied. LinUCB and Random were not included because the algorithms
are much simpler and thus faster (but also perform much
worse). Experiments were conducted on a computer with 16 cores (Intel
Xeon CPU L5520 @ 2.27GHz) and 32GB main memory. No multi-threading or
GP-GPU were used in the comparisons. The programming language R is
used to implement all the six algorithms.

From the results in Figure~\ref{fig:Efficiency-comparison-between}, we
can see that time consumed by both MCMC and variational inference
grows linearly with the training set size. However, variational
inference is more than 100 times faster than the MCMC, and
significantly faster than the L-BFGS-B algorithm. Comparing the
variational inference algorithm with two factors and it with three
factors, we find that adding another factor to the approximate
Bayesian model only slightly slows down the variational inference
algorithm. Moreover, when the sample size is less than 1000, the
variational inference algorithm can finish in 2 seconds, which makes
online updating practical and meets the user requirement
well. Implementing the algorithms in more efficient languages like
C/C++ can result in even better efficiency.

Time consumed in the prediction phase of the Bayesian methods is
larger than that of Greedy and LinUCB-based methods because of the
sampling process. However, for the two factors model Bayes-UCB-CN-V,
prediction can be accelerated significantly by the PRODCLIN algorithm
without sacrificing the accuracy~\cite{MacKinnon2007Distribution}. In
addition, since prediction for different songs is trivially
parallelizable, scaling variational inference to large music databases
should be easy.

\subsection{User study}

\label{sub:User-study}
\begin{figure}
\centering
\begin{minipage}{0.48\linewidth}
  \includegraphics[scale=0.16]{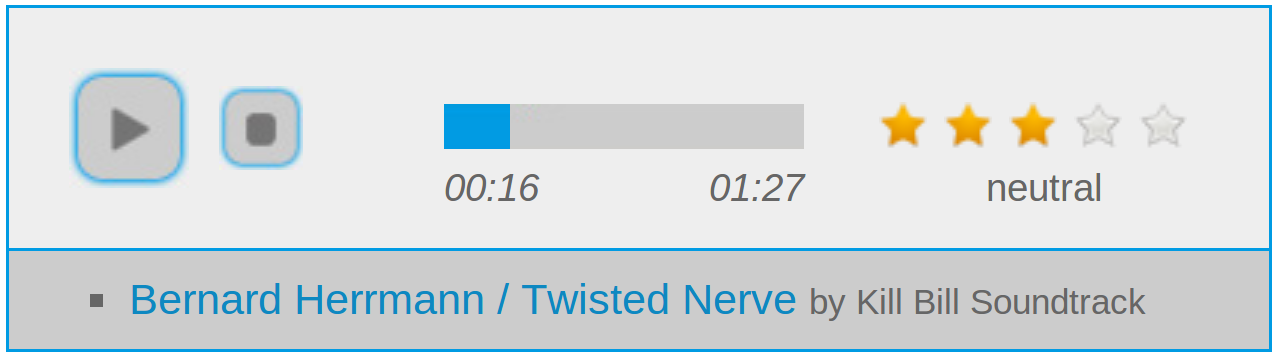}
  \captionof{figure}{User evaluation interface}
  \label{fig:Evaluation-interface}
\end{minipage}
\begin{minipage}{0.48\linewidth}
\centering
  \includegraphics[bb=0bp 0bp 504bp 445bp,clip,scale=0.40]{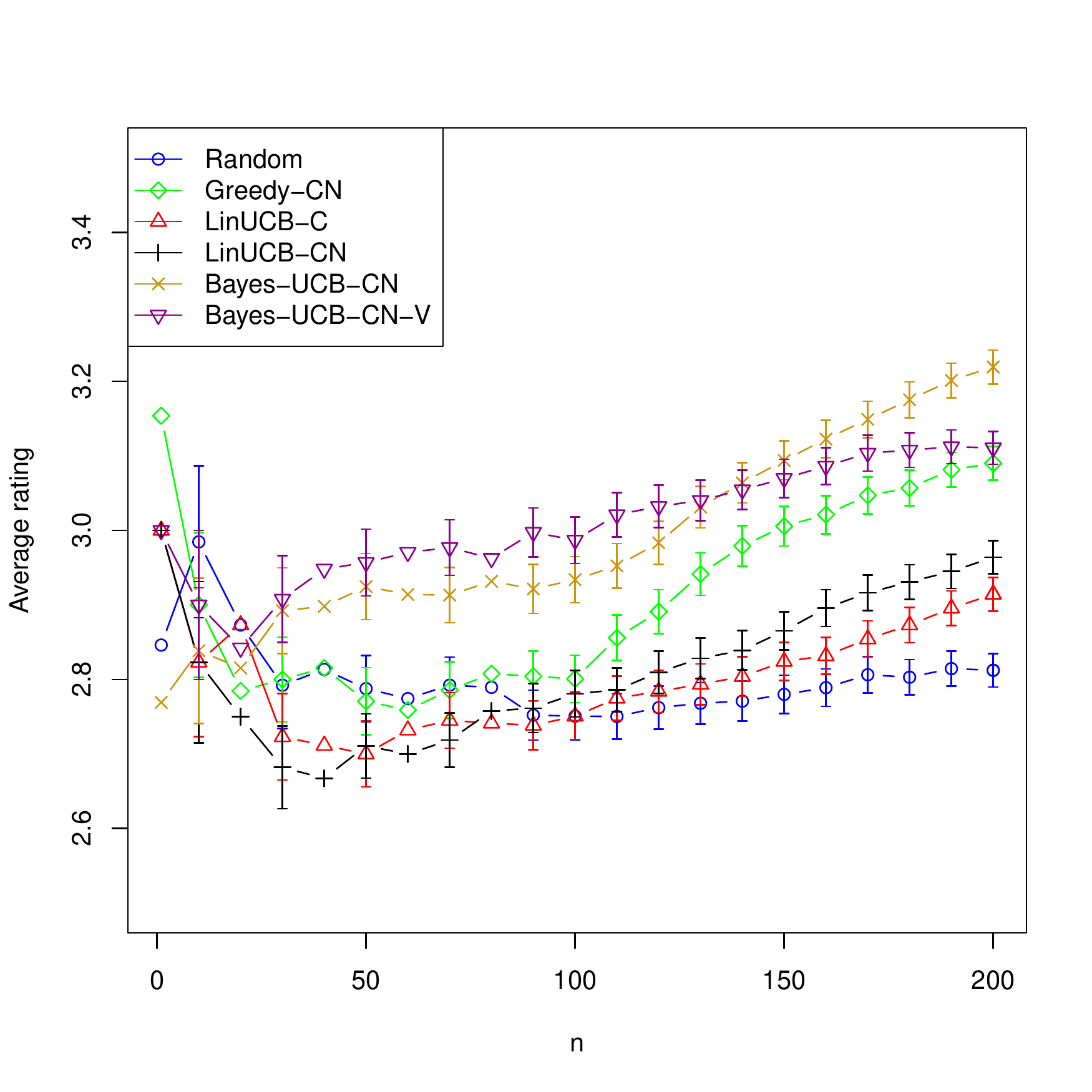}
  \vspace{-4mm}
  \captionof{figure}{Performance comparison in user study}
  \label{fig:Recommendation-Performance}
\end{minipage}
\end{figure}

Fifteen subjects (9 females and 6 males)
participated in the evaluation process. All are undergraduate students
with different majors and cultural backgrounds including Chinese,
Malay, Indian and Indonesian. All listen to music regularly (at least
3 hours per week).  Every subject was rewarded with a small token
payment for their effort and time. For each of the $6$ algorithms, a
subject evaluated $200$ recommendations, a number more than sufficient
to cover the cold-start stage. Every recommended song was listened to
for at least $30$ seconds (except when the subject was very familiar
with the song \textit{a priori}) and rated based on a $5$-point
Likert-scale as before. Subjects were required to rest for at least
$4$ minutes after listening to 20 songs to ensure the quality of the
ratings and simulate recommendation sessions. To minimize the
carryover effect, subjects were not allowed to evaluate more than two
algorithms within one day, and there must be a gap of more than 6
hours between two algorithms. The user study lasted for one
week. Every subject spent more than 14 hours in total.  The dataset
will be released after the publication of this paper.  During the
evaluation, the recommendation models were updated immediately
whenever a new rating was obtained. The main interface used for
evaluation is shown as Figure~\ref{fig:Evaluation-interface}.

\subsubsection{The overall recommendation performance}

\label{sub:Recommendation-performance}Because the true model is not
known in user study, the regret used in simulations cannot be used
here. We thus choose average rating as the evaluation metric, which is
also popular in evaluations of RL
algorithms. Figure~\ref{fig:Recommendation-Performance} shows the
average ratings and standard errors of every algorithm from the
beginning to the $n$-th recommendation.

T-tests at different iterations show Bayes-UCB-CN outperforms
Greedy-CN since the $45$th iteration with $p$-values <
$0.039$. Bayes-UCB-CN-V outperforms Greedy-CN from the $42$th to the
$141$th iteration with $p$-values < $0.05$, and afterwards with
$p$-values < $0.1$. Bayes-UCB-CN and Greedy-CN share the same rating
model and the only difference between them is that Bayes-UCB-CN
balances exploration/exploitation while Greedy-CN only
exploits. Therefore, the improvement of Bayes-UCB-CN against Greedy-CN
is solely contributed by the exploration/exploitation tradeoff,
affirming its effectiveness.

\begin{figure}
\centering
\begin{minipage}[b]{0.38\textwidth}
  \centering
  \includegraphics[bb=10bp 18bp 365bp 302bp,scale=0.50]{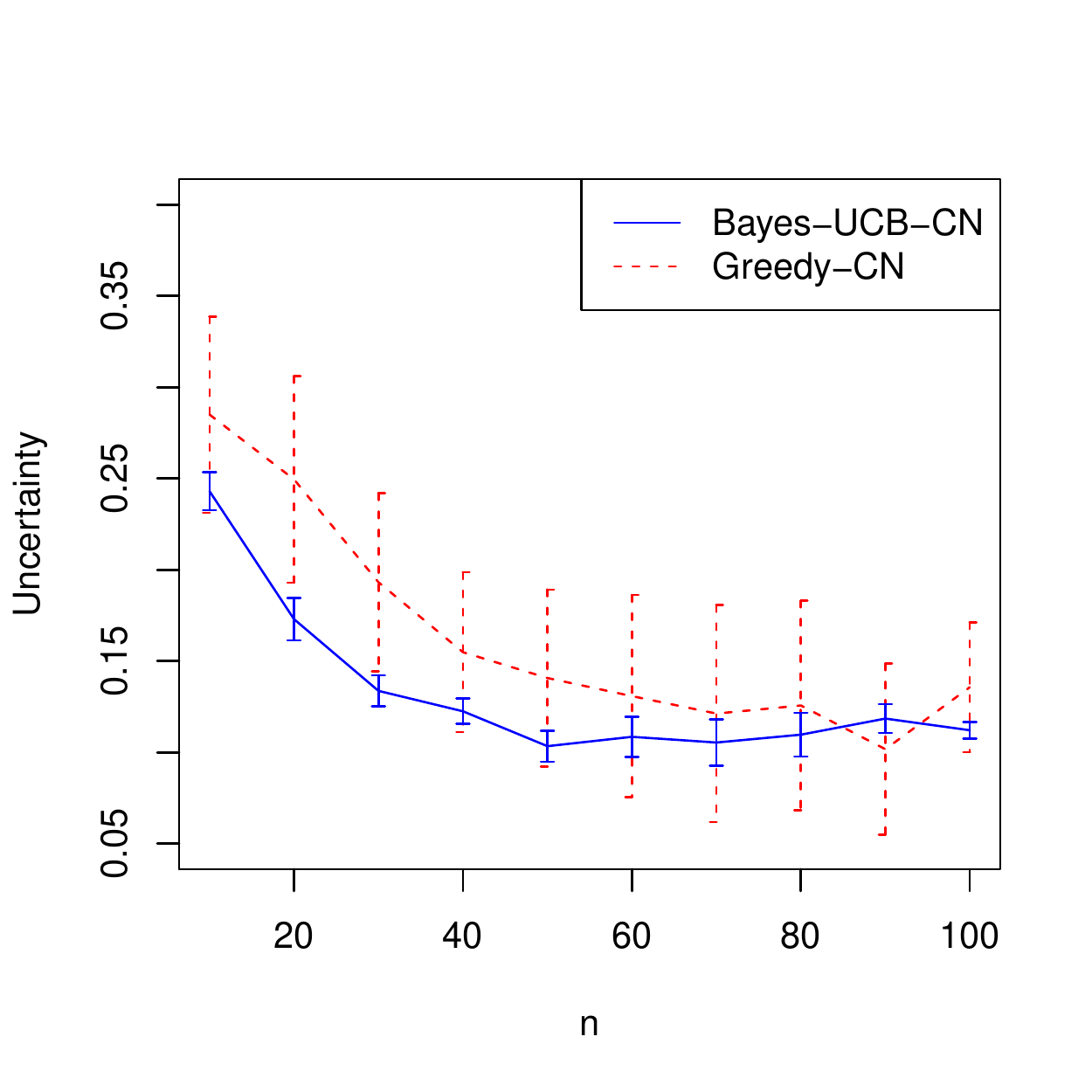}
  \captionof{figure}{Uncertainty}
  \label{fig:Uncertainty}
\end{minipage}
\hspace{3mm}
\begin{minipage}[b]{0.59\textwidth}
  \centering
  \includegraphics[bb=0bp 0bp 232bp 228bp,clip,scale=0.36]{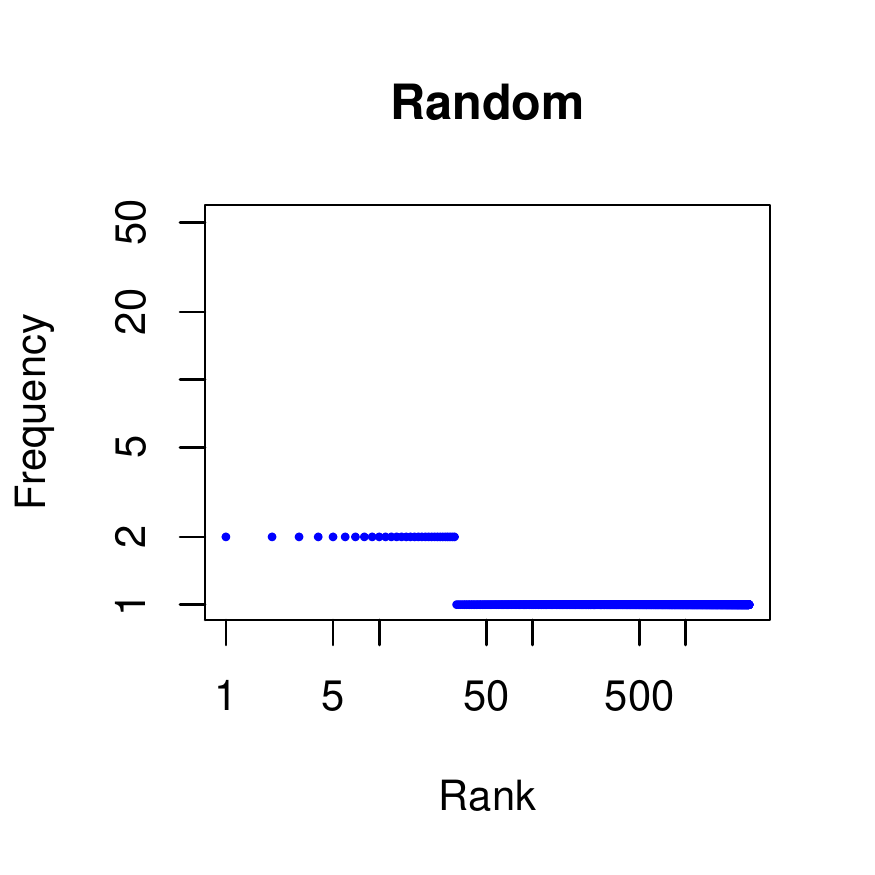}
  \includegraphics[bb=0bp 0bp 232bp 228bp,clip,scale=0.36]{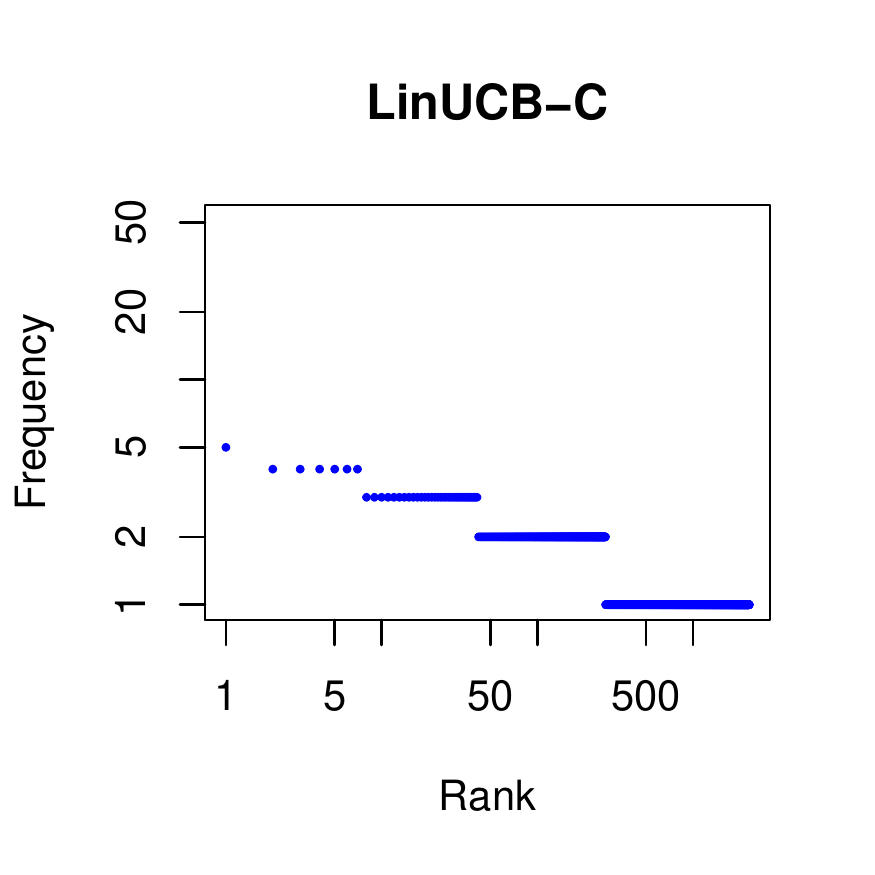}
  \includegraphics[bb=0bp 0bp 232bp 228bp,clip,scale=0.36]{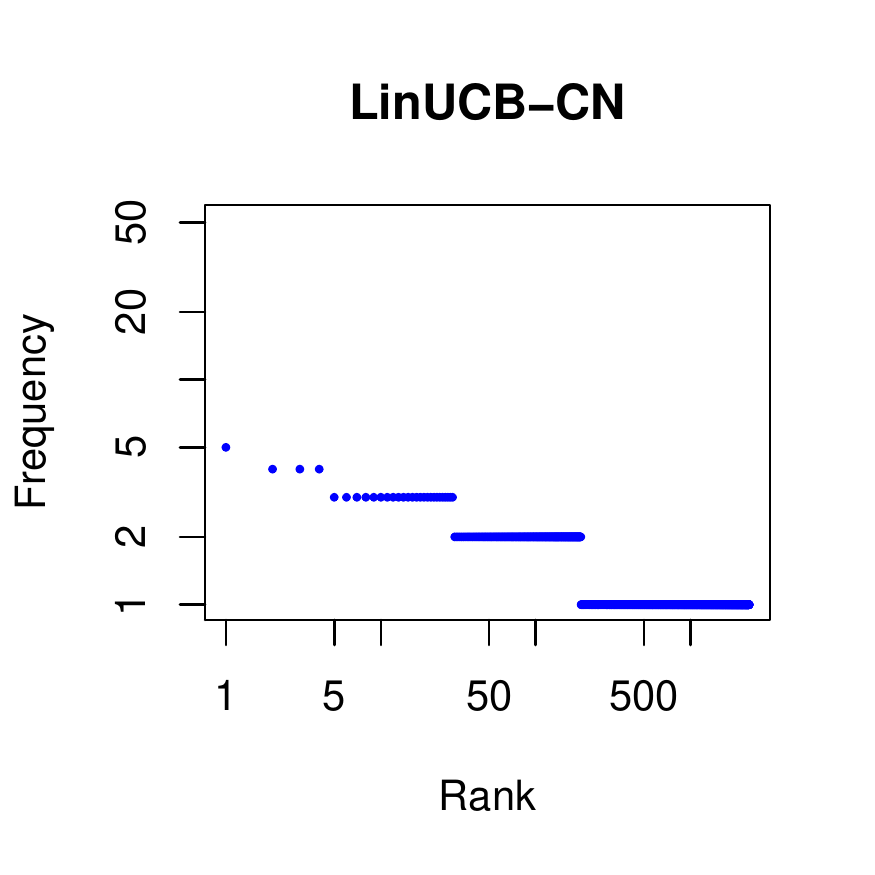}\par\vspace{-3mm}
  \includegraphics[bb=0bp 0bp 232bp 252bp,clip,scale=0.36]{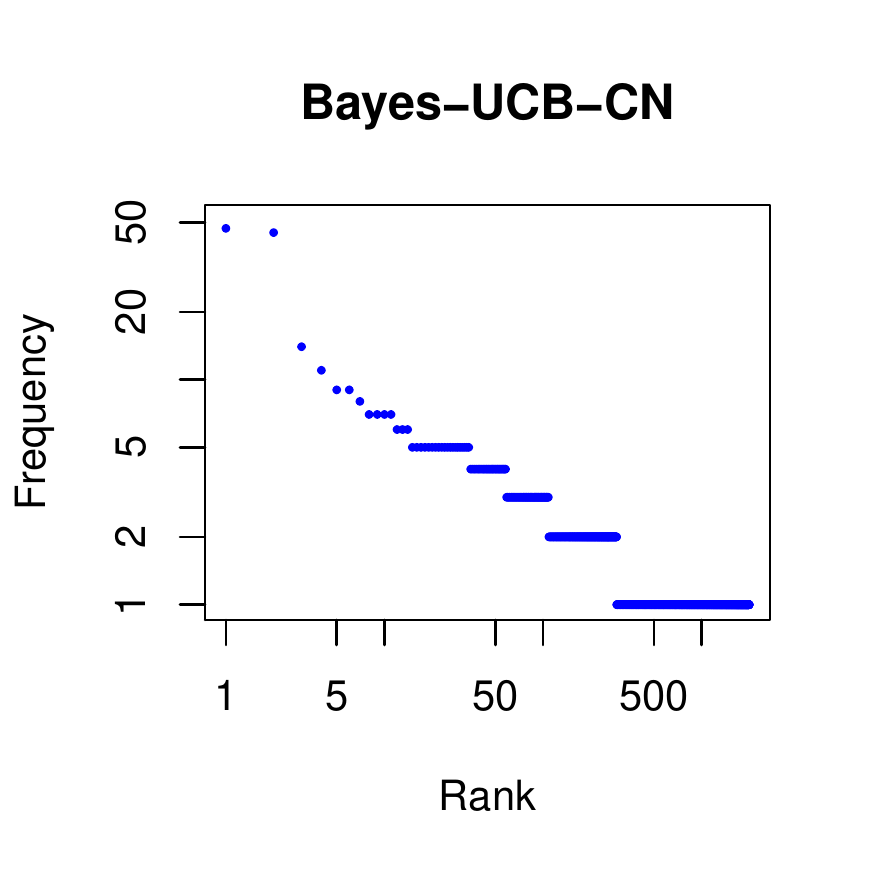}
  \includegraphics[bb=0bp 0bp 232bp 252bp,clip,scale=0.36]{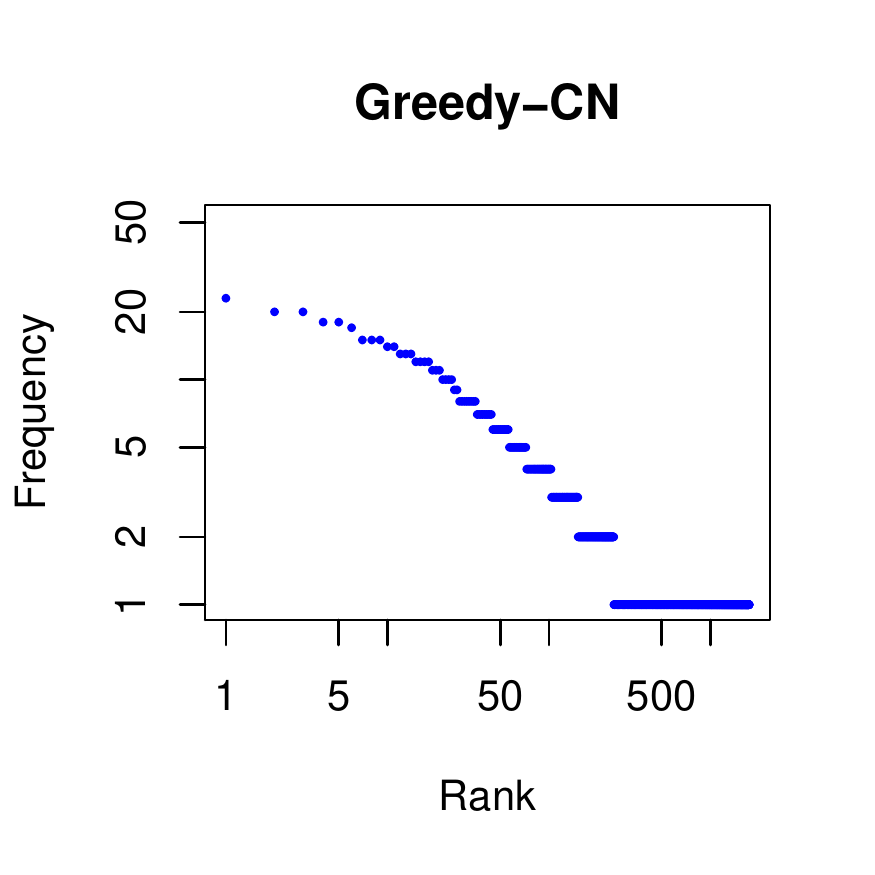}
  \includegraphics[bb=0bp 0bp 232bp 252bp,clip,scale=0.36]{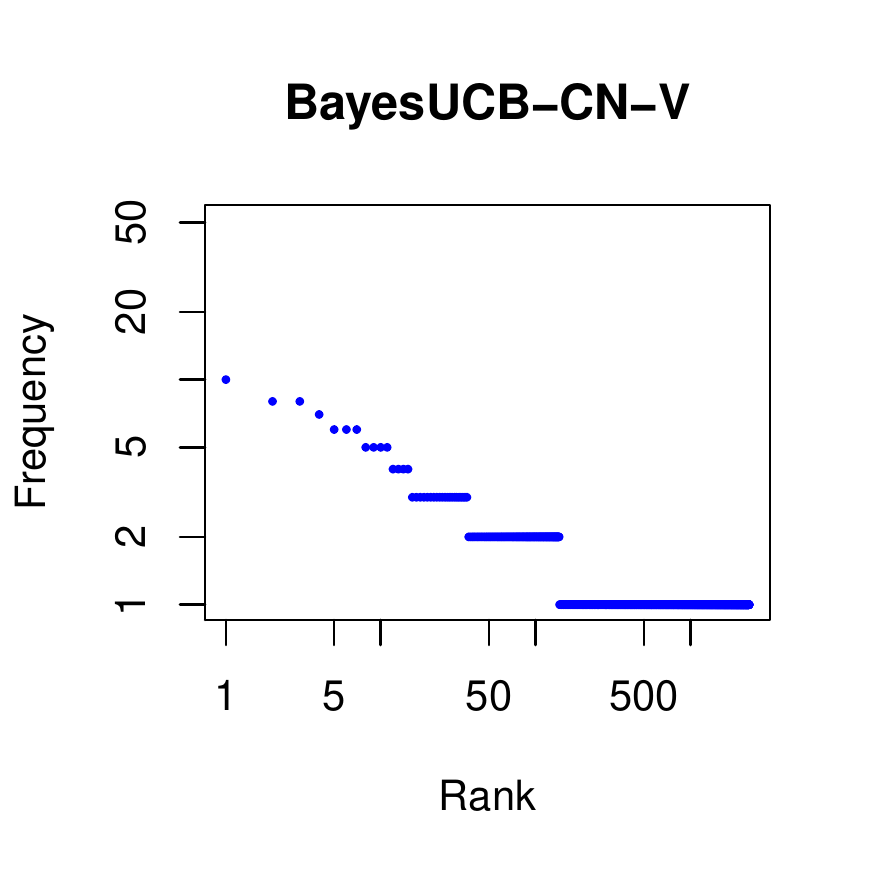}
  \vspace{-3mm}
  \captionof{figure}{Distributions of song repetition frequency}
  \label{fig:Distributions-of-repetition}
\end{minipage}
\end{figure}

More interestingly, when $n\leq100$ (cold-start stage) the differences
between Bayes-UCB-CN and Greedy-CN are even more significant. This is
because during the cold-start stage, the uncertainty is very high;
Bayes-UCB explores and thus reduces the uncertainty quickly while
Greedy-CN always exploits and thus cannot reduce the uncertainty as
efficiently as Bayes-UCB-CN. To verify this point, we first define a
metric for uncertainty as
\[
\mbox{uncertainty}=\frac{1}{|\mathcal{S}|}\sum_{k=1}^{|\mathcal{S}|}\mbox{SD}\left[p(U_{k}|\mathcal{D}_{n})\right]
\]
which is the mean of the standard deviations of all song's posterior
distributions $p(U_{k}|D_{n})$ estimated using the exact Bayesian
model. Larger standard deviation means larger uncertainty as
illustrated in Figure~\ref{fig:Uncertainty-in-recommendation.}. Given
the iteration $n$, we calculate an uncertainty measure based on each
user's recommendation history. The means and standard errors of the
uncertainties among all users at different iterations are shown in
Figure~\ref{fig:Uncertainty}.  When the number of training data points
increases, the uncertainty decreases. Also as expected, the
uncertainty of Bayes-UCB-CN decreases faster than Greedy-CN when $n$
is small, and later the two remain comparable because both have
obtained enough training samples to fully train the models. Therefore,
this verifies that our bandit approach handles uncertainty better
during the initial stage, and thus mitigate the cold-start problem.

Results in Figure~\ref{fig:Recommendation-Performance} also show that
all algorithms involving CN outperforms LinUCB-C, indicating that the
novelty factor of the rating model improves recommendation
performance. In addition, Bayes-UCB-CN outperforms LinUCB-CN
significantly, suggesting that multiplying $U_{c}$ and $U_{n}$
together works better than linearly combining them.

\subsubsection{Playlist generation\label{sub:Playlist-generation}}

As discussed in Section~\ref{sub:The-rating-model}, repeating songs
following the Zipf's law is important for playlist generation.
Therefore, we evaluated the playlists generated during the
recommendation process by examining the distribution of songs
repetition frequencies for every user. We generated the plots of the
distributions in the same way we generated
Figure~\ref{fig:Repetition-frequency-of} for the six algorithms. Ideal
algorithms should reproduce repetition distributions of
Figure~\ref{fig:Repetition-frequency-of}.

The results of the six algorithms are shown in
Figure~\ref{fig:Distributions-of-repetition}.  As we can see all
algorithms with $U_{c}$ and $U_{n}$ multiplied together
(i.e. Bayes-UCB-CN, Greedy-CN, BayesUCB-CN-V) reproduce the Zipf's law
pattern well, while the algorithms without $U_{c}$ (Random, LinUCB-C)
or with $U_{c}$ and $U_{n}$ added together (LinUCB-CN) do not. This
confirms that our model $U=U_{c}U_{n}$ can effectively reproduce the
Zipf's law distribution. Thus, we successfully modeled an important
part for combining music recommendation and playlist generation.

\subsubsection{Piecewise linear approximation}

\begin{figure}
\begin{centering}
\vspace{-8mm}
\subfloat{\includegraphics[bb=0bp 0bp 260bp 230bp,clip,scale=0.4]{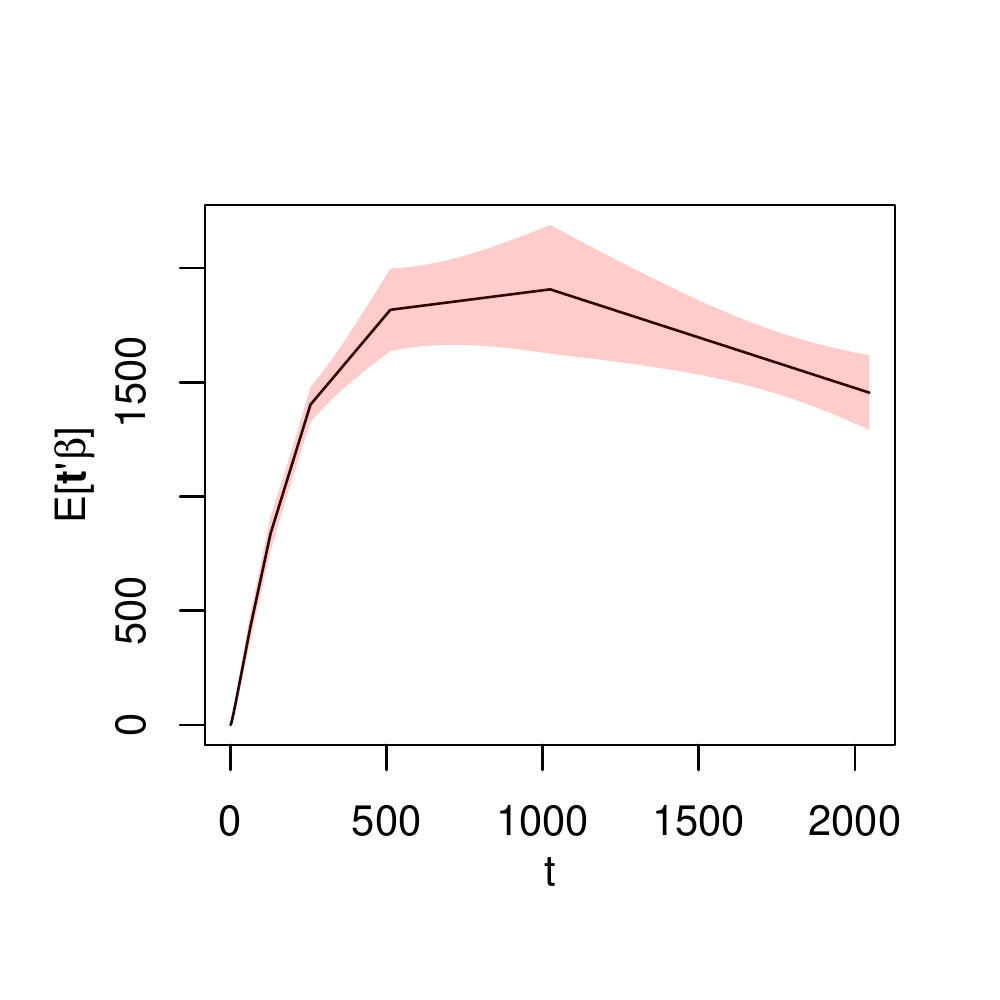}}\subfloat{\includegraphics[bb=0bp 0bp 260bp 230bp,clip,scale=0.4]{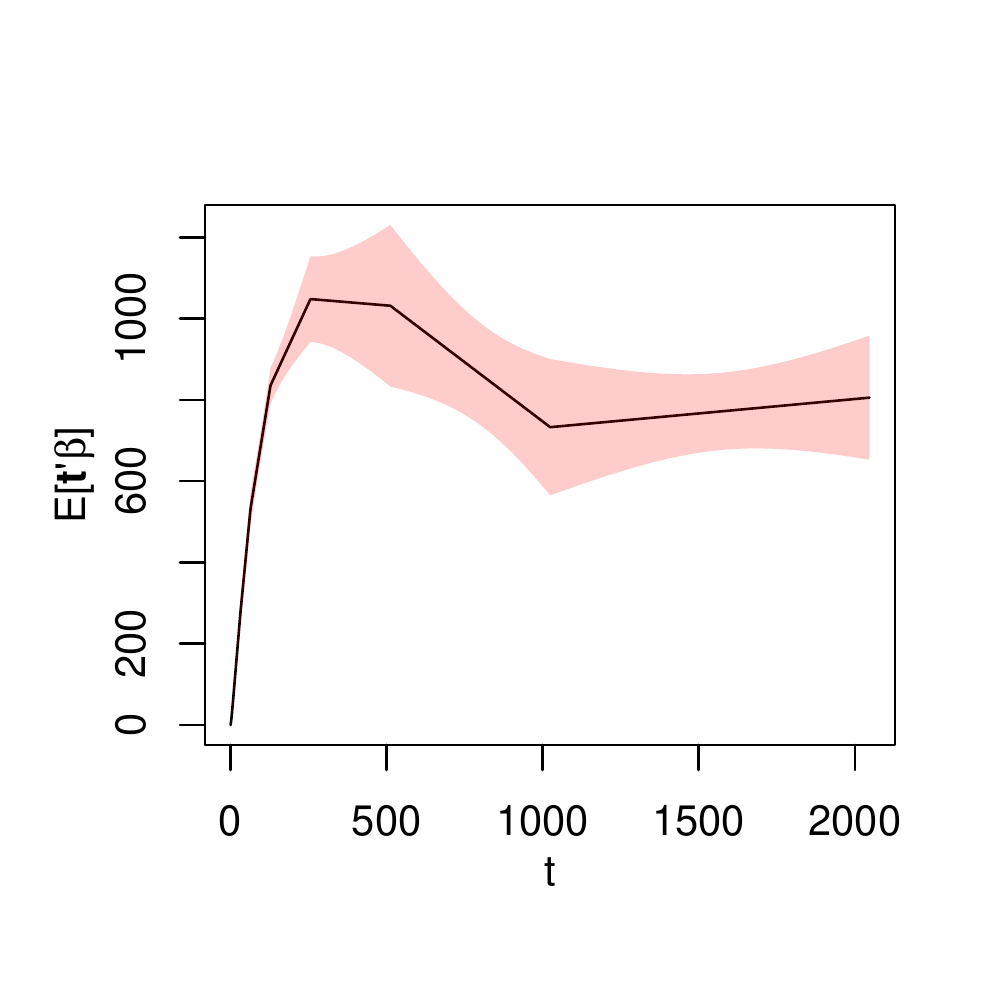}}
\subfloat{\includegraphics[bb=0bp 0bp 260bp 288bp,scale=0.4]{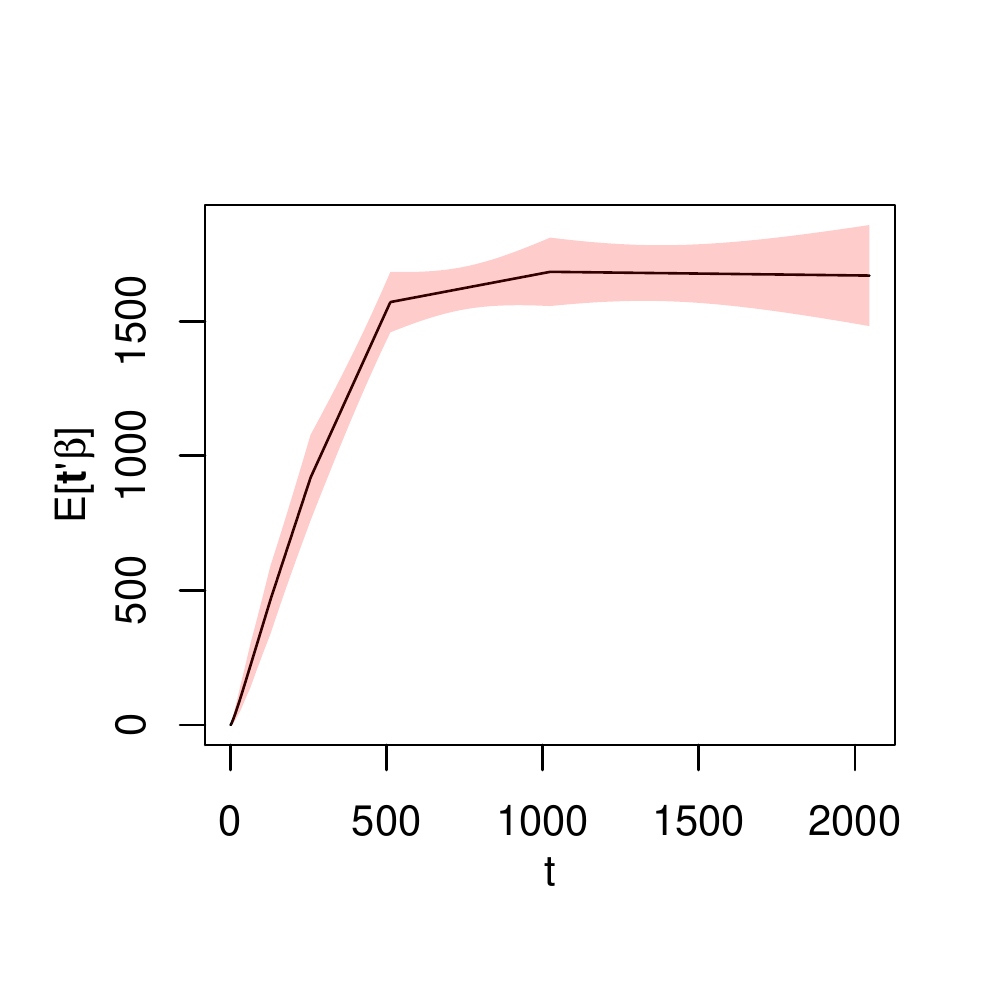}}\subfloat{\includegraphics[bb=0bp 0bp 260bp 288bp,scale=0.4]{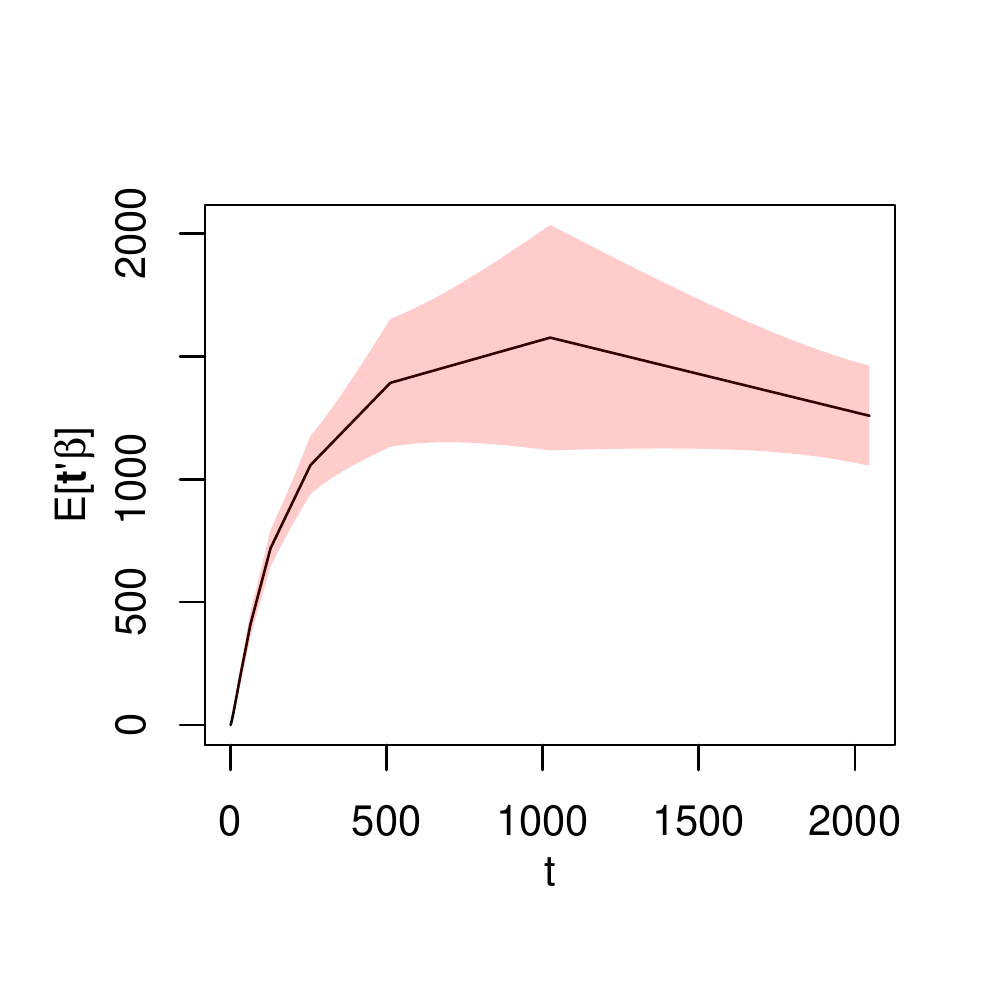}}
\par\end{centering}
\vspace{-5mm}
\caption{Four users' diversity factors learnt from the approximate Bayesian
model \label{fig:novelty}}
\end{figure}

In addition to the studies detailed above, the piecewise linear
approximation of the novelty model is tested again by randomly
selecting four users and showing in Figure~\ref{fig:novelty} their
novelty models learnt by Bayes-UCB-CN-V. Specifically, the posterior
distributions of $\bs{\beta}^{\prime}\mbf{t}$ for
$t\in(0,2^{11})$ are presented. Black lines represent the mean values
of $\bs{\beta}^{\prime}\mbf{t}$ and the red regions the
confidence bands of one standard deviation. The scale of
$\bs{\beta}^{\prime}\mbf{t}$ is not important because
$\bs{\beta}^{\prime}\mbf{t}$ is multiplied together with
the content factor, and any constant scaling of one factor can be
compensated by the scaling of the other one.  Comparing
Figure~\ref{fig:novelty} and Figure~\ref{fig:u2u3}, we can see that
the learnt piecewise linear novelty factors match our analytic form
$U_{n}$ well. This again confirms the accuracy of the piecewise linear
approximation.

\section{Discussion}
\label{discussion}

Exploring user preferences is a central issue for recommendation
systems, regardless of the specific media types. Under uncertainty,
the greedy approach usually results in suboptimal results, and
balancing exploration/exploitation is important. One successful
example of exploration/exploitation tradeoff is the news
recommender~\cite{LiLihong10Contextual}. Our work in this paper has
shown its effectiveness in music recommendation. Given that
uncertainty exists universally in all kinds of recommenders, it will
be interesting to examine its effectiveness in recommenders for other
media types such as video and image.

Also, our models and algorithms could be generalized to other
recommenders. First, the mathematical form of the approximate Bayesian
model is general enough to cover a family of rating functions that can
be factorized as the product of a few linear functions
(Section~\ref{sub:Integration-of-other-utility-functions}). Moreover,
we can often approximate nonlinear functions with linear ones. For
instance, we can use a feature mapping function $\phi(\mbf{x})$ and
make $U_c=\bs{\theta}^\prime\phi(\mbf{x})$ to capture the
non-linearity in our content model. Therefore, it will be interesting
to explore our approximate Bayesian model and the variational
inference algorithm in other recommendation systems. Second, the
proposed novelty model may not be suitable for movie recommendation
due to different consumption patterns in music and movie---users may
listen to their favorites songs for many times, but repetitions are
relatively rare for movies. However, the novelty model may suit
recommenders which repeat items (e.g. food or makeup
recommenders~\cite{Liu2013MM}). If their repetition patterns also
follow the Zipf's law, both the exact and approximate Bayesian models
can be used; otherwise, the approximate Bayesian model can be used at
least.

As for extensions of this work, the first interesting direction is to
model the correlations between different users to further reducing the
amount of exploration. This could be achieved by extending the
Bayesian models to hierarchical Bayesian models. Another interesting
direction is to consider more factors such as diversity, mood, and
genres to generate even better playlists, for the integration of
which, our approximate Bayesian model could be a good starting point.

\section{Conclusion}

\label{sec:Discussion-and-future} In this paper, we have described a
bandit approach to interactive music recommendation that balances
exploration and exploitation, mitigates the cold-start problem, and
improves recommendation performance. We have also described a rating
model including music audio content and novelty to integrate music
recommendation and playlist generation. To jointly learn the
parameters of the rating model, a Bayesian regression model together
with a MCMC inference procedure were developed. To make the Bayesian
inference efficient enough for online updating and generalize the
model for more factors such as diversity, a piecewise linear
approximate Bayesian regression model and a variational inference
algorithm were built.  The results from simulation demonstrate that
our models and algorithms are accurate and highly efficient. User
study results show that (1) the bandit approach mitigates the
cold-start problem and improves recommendation performance, and (2)
the novelty model together with the content model capture the Zipf's
law of repetitions in recommendations.

\begin{scriptsize}
\bibliographystyle{acmlarge}
\bibliography{wangxinxi}
\end{scriptsize}

\appendix
\section*{APPENDIX}
\setcounter{section}{1}
\label{sec:variational_inference}

The following is the variational lower bound, where $\psi(\cdot)$
is the digamma function. {\footnotesize 
\begin{align*}
\mathcal{L} & =\bb{E}[\ln(\mathcal{D},\tau,\bs{\theta},\bs{\beta})]-\bb{E}[\ln q(\bs{\theta},\tau,\bs{\beta})]\\
  & =\bb{E}\left[\ln p(\tau)\right]+\bb{E}\left[\ln p(\bs{\theta}|\tau)\right]+\bb{E}\left[\ln p(\bs{\beta}|\tau)\right]
 +\sum_{i=1}^{N}\bb{E}\left[\ln p(r_{i}|\mbf{x}_{i},\mbf{t}_{i},\bs{\theta},\bs{\beta},\tau)\right]
 -\bb{E}\left[\ln q(\bs{\theta})\right]-\bb{E}\left[\ln q(\bs{\beta})\right]-\bb{E}\left[\ln q(\tau)\right]\\
 & =a_{0}\ln b_{0}+(a_{0}-1)\left[\psi(a_{N})-\ln b_{N}\right]-b_{0}\frac{a_{N}}{b_{N}}
  -\frac{p}{2}\ln(2\pi)-\frac{1}{2}\ln\left|\mbf{D}_{0}\right|+\frac{p}{2}\left(\psi(a_{N})-\ln(b_{N})\right)\\
 & -\frac{a_{N}}{2b_{N}}\left[\tr(\mbf{D}_{0}\bs{\Lambda}_{\bs{\theta}N}^{-1})+(\bs{\mu}_{\bs{\theta}0}-\bb{E}\left[\bs{\theta}\right])^{\prime}\mbf{D}_{0}^{-1}(\bs{\mu}_{\bs{\theta}0}-\bb{E}\left[\bs{\theta}\right])\right]
  -\frac{K}{2}\ln(2\pi)-\frac{1}{2}\ln\left|\mbf{E}_{0}\right|+\frac{K}{2}\left(\psi(a_{N})-\ln(b_{N})\right)\\
 & -\frac{a_{N}}{2b_{N}}\left[\tr(\mbf{E}_{0}\bs{\Lambda}_{\bs{\beta}N}^{-1})+(\bs{\mu}_{\bs{\beta}0}-\bb{E}\left[\bs{\beta}\right])^{\prime}\mbf{E}_{0}^{-1}(\bs{\mu}_{\bs{\beta}0}-\bb{E}\left[\bs{\beta}\right])\right]
  -\frac{1}{2}\ln(2\pi)+\frac{1}{2}\left(\psi(a_{N})-\ln b_{N}\right)\\
 & -\frac{a_{N}}{2b_{N}}\sum_{i=1}^{N}\left(r_{i}^{2}+\mbf{x}_{i}^{\prime}\bb{E}\left[\bs{\theta}\bs{\theta}^{\prime}\right]\mbf{x}_{i}\mbf{t}_{i}^{\prime}\bb{E}\left[\bs{\beta}\bs{\beta}^{\prime}\right]\mbf{t}_{i}\right)
 +\frac{a_{N}}{b_{N}}\sum_{i=1}^{N}r_{i}\mbf{x}_{i}^{\prime}\bb{E}\left[\bs{\theta}\right]\mbf{t}_{i}^{\prime}\bb{E}\left[\bs{\beta}\right] +\frac{K}{2}\left[1+\ln(2\pi)\right]+\frac{1}{2}\ln\left|\bs{\Lambda}_{\bs{\beta}N}^{-1}\right|\\
& +\frac{p}{2}\left[1+\ln(2\pi)\right]
  +\frac{1}{2}\ln\left|\bs{\Lambda}_{\bs{\theta}N}^{-1}\right|-(a_{N}-1)\psi(a_{N})-\ln b_{N}+a_{N}
\end{align*}
}{\footnotesize \par}
The moments of $\bs{\theta}$, $\bs{\beta}$, and
$\tau$: {\footnotesize 
\[
\bb{E}\left[\bs{\beta}\bs{\beta}^{\prime}\right] =\bs{\Lambda}_{\beta N}^{-1}+\bb{E}[\bs{\beta}]\bb{E}[\bs{\beta}^{\prime}], \quad
\bb{E}[\bs{\beta}] =\bs{\Lambda}_{\bs{\beta}N}^{-1}\bs{\eta}_{\bs{\beta}N}, \quad
\bb{E}\left[\bs{\theta}\bs{\theta}^{\prime}\right]  =\bs{\Lambda}_{\bs{\theta}N}^{-1}+\bb{E}[\bs{\theta}]\bb{E}[\bs{\theta}^{\prime}], \quad
\bb{E}[\bs{\theta}] =\bs{\Lambda}_{\bs{\theta}N}^{-1}\bs{\eta}_{\bs{\theta}N}, \quad
\bb{E}[\tau] =\frac{a_{N}}{b_{N}}
\]
}%

\elecappendix
\newcommand{\D}{\mathcal{D}}
\newcommand{\bx}{\mathbf{x}}
\newcommand{\bt}{\mathbf{t}}
\newcommand{\btheta}{\boldsymbol{\theta}}
\newcommand{\bbeta}{\boldsymbol{\beta}}
\newcommand{\bmu}{\boldsymbol{\mu}}
\newcommand{\bLambda}{\boldsymbol{\Lambda}}
\newcommand{\bEta}{\boldsymbol{\eta}}
\newcommand{\Dzero}{\mathbf{D}_0}
\newcommand{\Ezero}{\mathbf{E}_0}

\section{Conditional distributions for the approximate Bayesian model}
Given $N$ training samples $\D=\{r_{i},\bx_{i},\bt_{i}\}_{i=1}^{N}$,
the conditional distribution $p(\btheta|\D,\tau,\bbeta)$ remains a
normal distribution as:
\begin{align*}
  p(\btheta|\D,\tau,\bbeta) & \propto p(\tau)p(\btheta|\tau)p(\bbeta|\tau)\prod_{i=1}^{N}p(r_i|\bx_i,\bt_i,\btheta,\bbeta,\tau)\\
  & \propto p(\btheta|\tau)\prod_{i=1}^{N}p(r_i|\bx_i,\bt_i,\btheta,\bbeta,\tau)\\
  & \propto\exp\left(-\frac{1}{2}\left(\btheta-\bmu_{\btheta0}\right)^{\prime}\left(\sigma^{2}\Dzero\right)^{-1}\left(\btheta-\bmu_{\btheta0}\right)\right)\exp\left(\sum_{i=1}^{N}-\frac{1}{2}\left(r_i-\bx_i^{\prime}\btheta \bt_i^{\prime}\bbeta\right)^{2}\left(\sigma^{2}\right)^{-1}\right)\\
  & \propto\exp\left[-\frac{1}{2}\btheta^{\prime}\left(\tau
      \Dzero^{-1}+\tau\sum_{i=1}^{N}\bx_i\bt_i^{\prime}\bbeta\bbeta^{\prime}\bt_i\bx_i^{\prime}\right)\btheta+\tau\left(\bmu_{\btheta0}^{\prime}\Dzero^{-1}+\sum_{i=1}^{N}r_i\bbeta^{\prime}\bt_i\bx_i^{\prime}\right)\btheta\right] \\
  & \propto\exp\left(-\frac{1}{2}\btheta^{\prime}\bLambda_{\btheta
    N}\btheta+\bEta_{\btheta_{N}}^{\prime}\btheta\right)
\end{align*}
where \begin{align*}
\bLambda_{\btheta N} & =\tau\left(\Dzero^{-1}+\sum_{i=1}^{N}\bx_i\bt_i^{\prime}\bbeta\bbeta^{\prime}\bt_i\bx_i^{\prime}\right)\\
\bEta_{\btheta N}^{\prime} & =\tau\left(\bmu_{\btheta0}^{\prime}\Dzero^{-1}+\sum_{i=1}^{N}r_i\bbeta^{\prime}\bt_i\bx_i^{\prime}\right)
\end{align*} 

Due to the symmetry between $\btheta$ and $\bbeta$, we can easily obtain
\begin{align*}
p(\bbeta|\D,\tau,\btheta) & \propto\exp\left(-\frac{1}{2}\bbeta^{\prime}\bLambda_{\bbeta N}\btheta+\bEta_{\bbeta N}^{\prime}\bbeta\right),
\end{align*}
where
\begin{align*}
\bLambda_{\bbeta N} & =\tau\left({\Ezero}^{-1}+\sum_{i=1}^{N}\bt_i\bx_i^{\prime}\btheta\btheta^{\prime}\bx_i\bt_i^{\prime}\right)\\
\bEta_{\bbeta N}^{\prime} & =\tau\left(\bmu_{\bbeta0}^{\prime}{\Ezero}^{-1}+\sum_{i=1}^{N}r_i\btheta^{\prime}\bx_i\bt_i^{\prime}\right).
\end{align*}

The conditional distribution $p(\tau|\D,\btheta,\bbeta)$ also remains
a Gamma distribution:
\begin{align*}
p(\tau|\D,\btheta,\bbeta) & \propto\tau^{a_{N}-1}\exp\left(-b_{N}\tau\right) \\
p(\tau|\D,\btheta,\bbeta) & \propto p(\tau)p(\btheta|\tau)p(\bbeta|\tau)\prod_{i=1}^{N}p(r_i|\bx_i,\bt_i,\btheta,\bbeta,\tau)\\
 & =b_{0}^{a_{0}}\frac{1}{\Gamma(a_{0})}\tau^{a_{0}-1}\exp(-b_{0}\tau)\times\\
 & \frac{1}{(2\pi)^{p/2}|\sigma^{2}{\Dzero}|^{1/2}}\exp\left(-\frac{1}{2}\left(\btheta-\bmu_{\btheta 0}\right)^{\prime}\left(\sigma^{2}{\Dzero}\right)^{-1}\left(\btheta-\bmu_{\btheta 0}\right)\right)\\
 & \times\frac{1}{(2\pi)^{K/2}|\sigma^{2}{\Ezero}|^{1/2}}\exp\left(-\frac{1}{2}\left(\bbeta-\bmu_{\bbeta0}\right)^{\prime}\left(\sigma^{2}{\Ezero}\right)^{-1}\left(\bbeta-\bmu_{\bbeta0}\right)\right)\\
 & \times\left(\frac{1}{(2\pi)^{1/2}|\sigma^{2}|^{1/2}}\right)^{N}\exp\left(\sum_{i=1}^{N}-\frac{1}{2}\left(r_i-\bx_i^{\prime}\btheta \bt_i^{\prime}\bbeta\right)^{2}\left(\sigma^{2}\right)^{-1}\right)\\
 & \propto\tau^{a_{N}-1}\exp\left(-b_{N}\tau\right)
\end{align*}
where $a_{N}$ and $b_{N}$ are the parameters of the Gamma
distribution, and they are
\begin{align*}
a_{N} & =\frac{p+K+N}{2}+a_{0}\\
b_{N} & =b_{0}+\frac{1}{2}\left(\btheta-\bmu_{\btheta 0}\right)^{\prime}{\Dzero}^{-1}\left(\btheta-\bmu_{\btheta 0}\right)+\frac{1}{2}\left(\bbeta-\bmu_{\bbeta0}\right)^{\prime}{\Ezero}^{-1}\left(\bbeta-\bmu_{\bbeta0}\right)\\
 & +\frac{1}{2}\sum_{i=1}^{N}\left(r_i-\bx_i^{\prime}\btheta \bt_i^{\prime}\bbeta\right)^{2}
\end{align*}

\section{Variational inference}

To calculate the joint posterior distribution
$p(\btheta,\tau,\bbeta|\D)$, we can use Gibbs sampling based on the
conditional distributions. However, this is slow too, and therefore,
we resort to variational inference (mean field approximation
specifically). We assume that $p(\btheta,\tau,\bbeta|\D)\approx
q(\btheta,\bbeta,\tau)=q(\btheta)q(\bbeta)q(\tau)$. In the restricted
distribution $q(\btheta,\bbeta,\tau)$, every variable is assumed
independent from the other variables. Because all the conditional
distributions $p(\btheta|\D,\tau,\bbeta)$,
$p(\tau|\D,\btheta,\bbeta)$, and $p(\bbeta|\D,\btheta,\bbeta)$ are in
the exponential families, their restricted distributions $q(\btheta)$,
$q(\bbeta)$,$q(\tau)$ lie in the same exponential families as their
conditional distributions.  We then obtain the restricted
distributions and update rules as in
Section~\ref{sub:Variational-Inference}.

The expectation of $b_{N}$ with respect to $q(\btheta)$ and
$q(\bbeta)$ might be a bit tricky to derive. We thus show it as the
following:
\begin{align*}
b_{N} & =b_{0}+\frac{1}{2}\mathbb{E}\left[\left(\btheta-\bmu_{\btheta 0}\right)^{\prime}{\Dzero}^{-1}\left(\btheta-\bmu_{\btheta 0}\right)\right]+\frac{1}{2}\mathbb{E}\left[\left(\bbeta-\bmu_{\bbeta0}\right)^{\prime}{\Ezero}^{-1}\left(\bbeta-\bmu_{\bbeta0}\right)\right]  +\frac{1}{2}\mathbb{E}\left[\sum_{i=1}^{N}\left(r_i-\bx_i^{\prime}\btheta \bt_i^{\prime}\bbeta\right)^{2}\right]\\
 & =b_{0}
  +\frac{1}{2}\left[\tr\left[{\Dzero}^{-1}\left(\mathbb{E}[\btheta\btheta^{\prime}]\right)\right]+\left(\bmu_{\btheta 0}^{\prime}-2\mathbb{E}[\btheta]^{\prime}\right){\Dzero}^{-1}\bmu_{\btheta 0}\right]\\
 & +\frac{1}{2}\left[\tr\left[{\Ezero}^{-1}\left(\mathbb{E}[\bbeta\bbeta^{\prime}]\right)\right]+\left(\bmu_{\bbeta0}^{\prime}-2\mathbb{E}[\bbeta]^{\prime}\right){\Ezero}^{-1}\bmu_{\bbeta0}\right]
  +\frac{1}{2}\sum_{i=1}^{N}\mathbb{E}\left[\left(r_i-\bx_i^{\prime}\btheta \bt_i^{\prime}\bbeta\right)^{2}\right].
\end{align*}
Since $\btheta$ and $\bbeta$ are assumed independent, we have 

\begin{align*}
\mathbb{E}\left[\left(r_i-\bx_i^{\prime}\btheta \bt_i^{\prime}\bbeta\right)^{2}\right] & =\mathbb{E}\left[r_i^{2}-2r_i\bx_i^{\prime}\btheta \bt_i^{\prime}\bbeta+\bx_i^{\prime}\btheta \bt_i^{\prime}\bbeta \bx_i^{\prime}\btheta \bt_i^{\prime}\bbeta\right]\\
 & =r_i^{2}-2r_i\bx_i^{\prime}\mathbb{E}[\btheta]\bt_i^{\prime}\mathbb{E}[\bbeta]+\bx_i^{\prime}\mathbb{E}[\btheta\btheta^{\prime}]\bx_i\bt_i^{\prime}\mathbb{E}[\bbeta\bbeta^{\prime}]\bt_i.
\end{align*}
Therefore $b_{N}$ can be calculated as

\begin{align*}
b_{N} & =b_{0} 
+\frac{1}{2}\left[\tr\left[{\Dzero}^{-1}\left(\mathbb{E}[\btheta\btheta^{\prime}]\right)\right]+\left(\bmu_{\btheta 0}^{\prime}-2\mathbb{E}[\btheta]^{\prime}\right){\Dzero}^{-1}\bmu_{\btheta 0}\right] \\
& + \frac{1}{2}\left[\tr\left[{\Ezero}^{-1}\left(\mathbb{E}[\bbeta\bbeta^{\prime}]\right)\right]+\left(\bmu_{\bbeta0}^{\prime}-2\mathbb{E}[\bbeta]^{\prime}\right){\Ezero}^{-1}\bmu_{\bbeta0}\right]\\
 & +\frac{1}{2}\left[\sum_{i=1}^{N}r_i^{2}-2r_i\bx_i^{\prime}\mathbb{E}[\btheta]\bt_i^{\prime}\mathbb{E}[\bbeta]+\bx_i^{\prime}\mathbb{E}[\btheta\btheta^{\prime}]\bx_i\bt_i^{\prime}\mathbb{E}[\bbeta\bbeta^{\prime}]\bt_i\right].
\end{align*}

\section{Variational lower bound}

It might be a bit tricky to derive 
\[ \mathbb{E}[\ln p(\btheta|\tau)]=\iint p(\btheta|\tau)q(\btheta)d\btheta q(\tau)d\tau \]
which is part of the lower bound $\mathcal{L}$. We assume that
$P=p(\btheta|\tau)$, and $Q=q(\btheta)$, and we have $\int
p(\btheta|\tau)q(\btheta)d\btheta=-H(Q,P)$, where $H(Q,P)$ is the cross
entropy between $Q$ and $P$. Given $Q$ and $P$ are multivariate normal
distributions, the KL-divergence between $Q$ and $P$ and the entropy
of $Q$ are
\begin{align*}
D_{KL}(Q\Vert P) & =\frac{1}{2}\left[\tr(\Sigma_{P}^{-1}\Sigma_{Q})+(\bmu_{P}-\bmu_{Q})^{\prime}\Sigma_{P}^{-1}(\bmu_{P}-\bmu_{Q})-\ln\frac{\left|\Sigma_{Q}\right|}{\left|\Sigma_{P}\right|}-p\right]\\
 & =\frac{1}{2}\left[\tr(\tau {\Dzero}\bLambda_{\btheta N}^{-1})+(\bmu_{\btheta 0}-\bmu_{\btheta N})^{\prime}\tau {\Dzero}^{-1}(\bmu_{\btheta 0}-\bmu_{\btheta N})-\ln\left|\bLambda_{\btheta N}^{-1}\right|+\ln\left|\frac{1}{\tau}{\Dzero}\right|-p\right]\\
H(Q) & =\frac{1}{2}(p+p\ln(2\pi)+\ln\left|\bLambda_{\btheta N}^{-1}\right|.
\end{align*}
Therefore
\begin{align*}
\int p(\btheta|\tau)q(\btheta)d\btheta & =-H(Q,P)\\
 & =-H(Q)-D_{KL}(Q\Vert P)\\
 & =-\frac{p}{2}\ln(2\pi)-\frac{1}{2}\ln\left|{\Dzero}\right|+\frac{p}{2}\ln\tau -\frac{1}{2}\left[\tr(\tau {\Dzero}\bLambda_{\btheta N}^{-1})+(\bmu_{\btheta 0}-\bmu_{\btheta N})^{\prime}\tau {\Dzero}^{-1}(\bmu_{\btheta 0}-\bmu_{\btheta N})\right],
\end{align*}
and
\begin{align}
\mathbb{E}[\ln p(\btheta|\tau)] & =\iint p(\btheta|\tau)q(\btheta)d\btheta q(\tau)d\tau\nonumber \\
 & =-\frac{p}{2}\ln(2\pi)-\frac{1}{2}\ln\left|{\Dzero}\right|+\frac{p}{2}\mathbb{E}[\ln\tau]\nonumber 
  -\frac{1}{2}\left[\tr({\Dzero}\bLambda_{\btheta N}^{-1})+(\bmu_{\btheta 0}-\bmu_{\btheta N})^{\prime}{\Dzero}^{-1}(\bmu_{\btheta 0}-\bmu_{\btheta N})\right]\mathbb{E}[\tau]\nonumber \\
 & =-\frac{p}{2}\ln(2\pi)-\frac{1}{2}\ln\left|{\Dzero}\right|+\frac{p}{2}\left(\psi(a_{N})-\ln(b_{N})\right)\nonumber \\
 & -\frac{a_{N}}{2b_{N}}\left[\tr({\Dzero}\bLambda_{\btheta N}^{-1})+(\bmu_{\btheta 0}-\bmu_{\btheta N})^{\prime}{\Dzero}^{-1}(\bmu_{\btheta 0}-\bmu_{\btheta N})\right]\nonumber
\end{align}
\end{document}